\documentclass[11pt]{article}
\usepackage{amsmath,amssymb,multirow,jheppub,graphicx, subfigure}
\usepackage{amsthm}
\usepackage{centernot}
\usepackage{enumitem}
\usepackage{color}
\allowdisplaybreaks[1]
\usepackage{tikz}
 \tikzset{node distance=2cm, auto}
\usepackage{bbm} 
\usepackage{youngtab}
\usepackage{slashed}
\usepackage{mathtools}
\usepackage{wrapfig}
\usepackage{ytableau}
\usepackage{xcolor}
\pagecolor{white}

\def\ZZ{\mathbb{Z}} 
\def\RR{\mathbb{R}}

\def\tfrac#1#2{{\textstyle{\frac{#1}{#2}}}}

\def\hat{\widehat}

\def\U{\textrm{U}}

\usepackage{verbatim}   
\usepackage[all]{xy}
\usepackage{bm}  
\def\Dslash{{\rlap{\raise 1pt \hbox{$\>/$}}D}}
\def\Pslash{{\rlap{\raise  1pt \hbox{$\>/$}}\,\partial}}

\newcommand{\be}{\begin{equation}}      
\newcommand{\ee}{\end{equation}}      
\newcommand{\beq}{\begin{eqnarray}}      
\newcommand{\eeq}{\end{eqnarray}}

\newcommand{\ZF}{\left(\mathbb{Z}_2\right)_F}

\newcommand{\UB}{U(1)^{(0)}_B}

\theoremstyle{definition}

\begin{document}

\title{Line operators, vortex statistics, and Higgs versus confinement dynamics}
\author[1]{Aleksey Cherman,}
\emailAdd{aleksey.cherman.physics@gmail.com}
\affiliation[1]{School of Physics and Astronomy, University of Minnesota, Minneapolis, MN 55455, USA}
\author[1,2]{Theodore Jacobson,}
\emailAdd{tjacobson@physics.ucla.edu}
\affiliation[2]{Mani L. Bhaumik Institute for Theoretical Physics, Department of Physics and Astronomy, University of California, Los Angeles, CA 90095, USA }
\author[3]{Srimoyee Sen,}
\emailAdd{srimoyee08@gmail.com}
\affiliation[3]{Department of Physics and Astronomy, Iowa State University, Ames, IA 50011, USA}
\author[4]{Laurence G. Yaffe}
\emailAdd{yaffe@phys.washington.edu}
\affiliation[4]{Department of Physics, University of Washington, Seattle, WA 98195, USA}

\abstract{We study a $2{+}1$D lattice gauge theory with
fundamental representation scalar fields which has both Higgs and confining regimes
with a spontaneously-broken $U(1)$ $0$-form symmetry.  We show that the
Higgs and confining regimes may be distinguished by a natural
gauge invariant observable: the phase $\Omega$ of a correlation function of a vortex line 
operator linking with an electric Wilson line. We employ dualities and
strong coupling expansions to analytically explore parameter regimes which were inaccessible in previous continuum calculations, and discuss possible implications for the phase diagram. 
}

\maketitle

\section{Introduction and summary}
Gauge theories with fundamental representation matter fields can exhibit both
confining and Higgs regimes, but it can be very difficult to construct useful
gauge invariant observables that sharply distinguish these two regimes.%
\footnote{See, for example, 
Refs.~\cite{Fredenhagen:1985ft,Greensite:2017ajx,Greensite:2018mhh,
Greensite:2020nhg,Greensite:2021fyi,Verresen:2022mcr,Thorngren:2023ple}.
} 
In these theories
it is generally expected that if all global symmetries have identical
realizations in both Higgs and confining regimes
then the two regimes need not be separated by a phase boundary
and that  there is no sharp physical distinction between them.%
\footnote
    {%
    This is provably true only in certain specific examples
    without global symmetries
    \cite{Fradkin:1978dv}.
    See also Refs.~\cite{Elitzur:1975im,
    Osterwalder:1977pc,
    Banks:1979fi,
    tHooft:1979yoe,
    Frohlich:1980gj,
    Frohlich:1981yi}.
    }

In this paper, which revisits and clarifies arguments originally
developed in
Refs.~\cite{Cherman:2020hbe,Cherman:2018jir},
we explore the distinction
between Higgs and confining regimes in gauge theories with
global symmetries whose realizations remain unaltered between the two regimes. However, the presence of these global symmetries opens up the possibility to evade the continuity
scenario of Fradkin and Shenker~\cite{Fradkin:1978dv}.%
\footnote
    {%
    Refs.~\cite{Verresen:2022mcr,Thorngren:2023ple} argue that in some gauge theories with fundamental representation matter, the existence of an emergent ``magnetic" global symmetry can be leveraged to distinguish the Higgs and confining regimes. In accordance with Fradkin-Shenker continuity, this does not necessarily imply a bulk phase boundary separating the two regimes, since the magnetic symmetry is not exact. 
    }
While the realization of the global symmetry is identical in the Higgs
and confining regimes in the system we study, the symmetry structure
will nevertheless allow us to define a natural, but non-local,
order parameter that distinguishes the two regimes. However, truly demonstrating the existence of a Higgs-confinement
phase boundary in the present setting requires either finding new
theoretical insights and/or performing careful numerical lattice simulations. The tools and analytic calculations presented in the current paper set the groundwork for future numerical investigations of the full phase diagram guided by the sharp change in behavior of our proposed order parameter. 

To review the basic questions motivating this paper in a simple
setting, consider first a $U(1)$ gauge theory in $2{+}1$ spacetime dimensions
coupled to a unit-charge Higgs field $\phi$. Suppose that the UV completion of
this Abelian-Higgs model admits finite-action monopole-instanton field
configurations, so that there is no nontrivial magnetic symmetry which would act
on monopole operators (which are local in $2{+}1$ dimensions).
When the Higgs field has a large positive mass-squared,
one can integrate it out and the long distance dynamics of local operators
is well-described by the Polyakov model, which is famously confining and
features a mass gap for the photon~\cite{Polyakov:1976fu}. It is natural to
describe the regime where $\phi$ is heavy as a confining regime in the usual
heuristic sense, despite the fact that sufficiently long confining flux tubes
will eventually break due to pair-production of dynamical $\phi$ particles.
In contrast, if $\phi$ condenses then one enters a Higgs regime where,
at least deep in the semiclassical domain,
it is clear that the gauge boson gets a mass related to the
``expectation value'' of $\phi$
(using perturbative gauge-dependent language).
These Higgs and confining regimes are not distinguished by the expectation value
of any local gauge invariant operator.
Both regimes are gapped, and in neither regime is any global symmetry
spontaneously broken.
Consequently, one may expect that they are part of the same phase.
This is corroborated by the
classic result of Fradkin and Shenker exhibiting a smooth path between the Higgs
and confining regimes of certain lattice gauge theories with fundamental representation 
matter and no global symmetries~\cite{Fradkin:1978dv}.

The situation can change dramatically in systems with global symmetries. One
example is furnished by a variant of the $2{+}1$D Abelian-Higgs model discussed
above but \emph{without} finite-action monopole-instantons, so that there is an
exact $U(1)_m$ magnetic global symmetry. The `condensing' unit-charge Higgs
field $\phi$ is neutral under the magnetic symmetry.
Nonetheless, the realization of $U(1)_m$ changes as one dials
the mass-squared parameter of $\phi$.
In the Higgs phase the $U(1)_m$ symmetry is unbroken,
while in the phase where $\phi$ is heavy the $U(1)_m$ symmetry
is spontaneously broken.
The phase with spontaneously broken $U(1)_m$ can be called confining,
as it features a logarithmic attractive
potential between probe electric charges (with opposite charges)
for distances up to the
`string-breaking' length scale.%
\footnote
    {%
    In Ref.~\cite{Cherman:2023xok} two of
    us recently argued that the 
    spontaneous breaking of $U(1)_m$ symmetry
    in the confining regime can be understood
    as a consequence of an emergent $1$-form symmetry which has
    an 't Hooft anomaly with the $U(1)_m$ symmetry.
    See also Ref.~\cite{Delacretaz:2019brr} for the related
    particle-vortex dual perspective.
    }
Therefore, global symmetry considerations
imply that the Higgs and confining regimes of this model are distinct phases. 

In this paper our interest lies in gauge theories where the Higgs fields are
charged under some global symmetry $G$.
In such theories, there can be an obvious distinction between
Higgs and confining regimes if the global symmetry $G$ is spontaneously broken
in the Higgs regime, but unbroken in the confining regime.
If so, then the Higgs and confining regimes must be distinct
phases of matter, separated by a phase boundary.
We focus on a subtler question: can one distinguish
Higgs and confining regimes when the realization of $G$ is the \emph{same}
in both regimes?
We will be particularly interested in distinguishing the Higgs and
confining regimes when the global symmetry $G$ is spontaneously broken. 

Apart from its intrinsic theoretical interest, this question is also motivated
by long-standing puzzles concerning the behavior of QCD as a function of
density.
Low temperature finite-density QCD has a spontaneously broken
$U(1)_B = U(1)/\mathbb{Z}_{N_c}$ baryon-number global symmetry (with $N_c=3$),
indicating that bulk nuclear matter is a superfluid.
As the baryon chemical potential is increased, it is natural to regard
dense QCD as evolving from a ``confining'' regime in which bound
nucleons provide the natural description,
to a ``Higgs'' regime in which, using gauge-variant language,
composite scalar di-quark operators condense.%
\footnote
    {%
    This asymptotically high density phase is termed
    a ``color superconductor'' in many papers.
    }
It is a long-standing open problem to determine whether these regimes
are sharply distinguishable from each other and separated by a
thermodynamic phase transition.
(See, e.g.,
Ref.~\cite{Alford:2007xm} for a review and references on the high-density
phase of QCD.)
In particular, Sch\"afer and Wilczek observed
that the Higgs and confining regimes of cold dense QCD,
near the $SU(3)$ flavor symmetric point,
have the same realization of all conventional global symmetries.
This motivated their well-known conjecture that the two regimes
are smoothly connected \cite{Schafer:1998ef}.
Their arguments relied on the traditional Landau
paradigm for distinguishing phases of matter based on realizations of
($0$-form) symmetries,
as well as the compatible nature of the spectrum of low-lying excitations,
and follow the general expectation of continuity based on
the classic work of Fradkin and Shenker~\cite{Fradkin:1978dv}.

Over the last several decades it has become clear that there
can be genuinely distinct phases of matter that are not visible within
the Landau paradigm,
see e.g.,~Refs.~\cite{Wen:2012hm,RevModPhys.89.041004,RevModPhys.88.035005,senthil:SPT_review,Gaiotto:2014kfa}.
This can, for example,
occur when differing phases of matter are distinguished by
intrinsically non-local observables, rather than the local order
parameters which play the starring role in the Landau-Ginzburg approach.
With this more general perspective in mind,
it is thus natural to ask whether the Higgs and confining
regimes of dense QCD can be sharply distinguished.
But given the complicated nature of QCD, it is
helpful to first examine, carefully, analogs of this
question in simpler settings.%
\footnote{For recent progress on distinguishing the Higgs and confining regimes
of dense QCD, see Ref.~\cite{Dumitrescu:2023hbe} which argued that the two
regimes are separated by a non-Landau phase transition associated with a jump in
the quantized gravitational theta term. This jump is associated with a phase
boundary in the time-reversal symmetric locus of parameter space, and is
directly related to the pattern of quark pairing in the high density phase while
being insensitive to the explicit breaking of baryon number symmetry. In
contrast, the type of transitions we contemplate in this paper would also occur
in bosonic theories and, given the role played by superfluid vortices, rely on a
spontaneously broken $U(1)$ symmetry. }

\begin{figure}
    \vspace*{-1cm}
    \centering
    \includegraphics[width=0.7\textwidth]{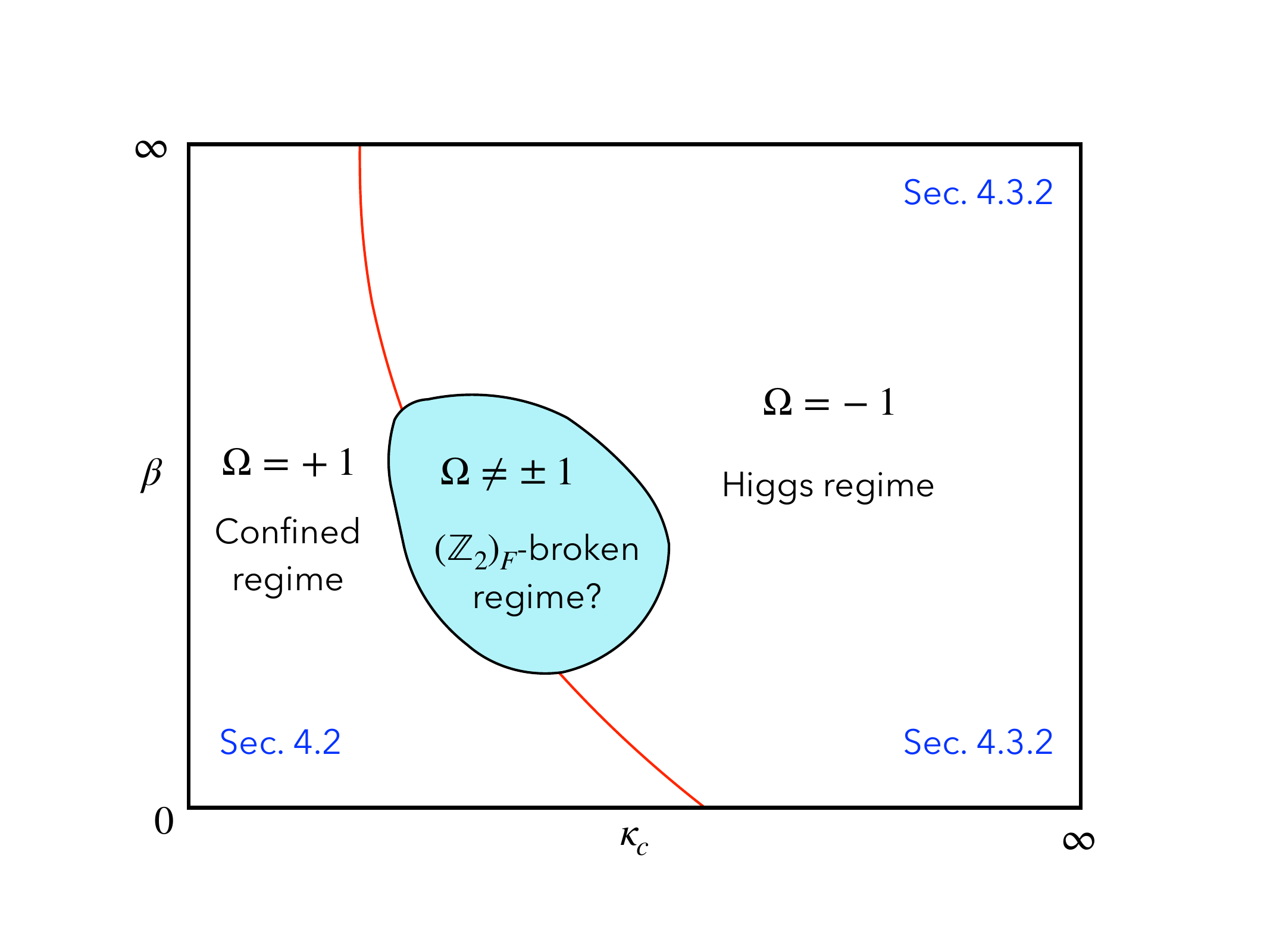}
    \label{fig:phase_diag}
    \caption
	{%
	[Color online]
	We study a gauge theory with global 
	$\UB$ and $\ZF$ $0$-form symmetries, and a
	$U(1)^{(1)}$ $1$-form symmetry.
	Due to a mixed 't Hooft anomaly involving $\UB$ and $U(1)^{(1)}$,
	the $\UB$ symmetry is always spontaneously broken.
	The $\ZF$ symmetry is unbroken in regimes connected to
	controllable limits,
	and leads to a quantization of the Aharonov-Bohm phase
	$\Omega = \pm 1$ acquired by electrically-charged probe particles
	when they encircle vortices.
	The value of this phase thus
	distinguishes the Higgs and confining regimes, as illustrated in the
	sketch.
	The axes show parameters of our lattice
	discretization of the theory, defined in Sec.~\ref{sec:lattice},
	with $\beta$ controlling a lattice gauge coupling and
	$\kappa_c$ a scalar field mass-squared.
	Indicated in the corners of the sketch are the
	sections in this paper where we evaluate the phase $\Omega$.
	As shown, the strongly-coupled region of parameter space
	may contain a region in which the
	$\ZF$ symmetry is spontaneously broken.
	}
\end{figure}

In this paper, following our earlier work~\cite{Cherman:2020hbe},
we will focus on a $U(1)$ gauge theory in $2{+}1$ dimensions. The gauge
theory we study has $\UB$ and $\ZF^{(0)}$ $0$-form symmetries, as well as
a $U(1)^{(1)}$ $1$-form symmetry which results from eliminating all dynamical vortices.%
\footnote
    {%
    Here and henceforth
    we use the notation $G^{(p)}$ to
    denote a $p$-form symmetry based on
    a group $G$.
    }
As will be discussed in Sec.~\ref{sec:vortex_operators},
the $\UB$ and $U(1)^{(1)}$ symmetries have a mixed 't Hooft
anomaly causing the $\UB$ symmetry to
always be spontaneously broken,
while the $U(1)^{(1)}$ symmetry remains unbroken.
As in dense QCD, this field
theory features Higgs and confining regimes, all within a gapless $\UB$-broken
phase. Though all conventional and generalized symmetries have identical
realizations in these two regimes, we argue that they are differentiated by the
values of the Aharonov-Bohm phase $\Omega$ of electric probe particles
encircling global vortices.
Physically, this phase appears if the vortices carry magnetic flux
in their cores.
With insights from a lattice regularization, we write $\Omega$
in terms of a correlation function 
of a Wilson loop $W(C)$ linked with a loop operator $V(\tilde\Gamma)$ which inserts a vortex,%
\footnote
    {%
    The minimal separation between the two loops must be large
    compared to the vortex core size. See also Footnote~\ref{fn:core_size} below.  
    }
\begin{equation}
    \label{eq:order_parameter}
    \Omega = \exp\left[ i
    \arg\, \langle V(\tilde\Gamma) W(C) \rangle \right].
\end{equation} 
Unbroken $\ZF^{(0)}$ symmetry implies that the phase $\Omega$ must be real. We
will analytically compute the $\Omega$  correlation function
\eqref{eq:order_parameter} in tractable portions of the parameter space of our
theory and show that $\Omega$ takes distinct values, namely $\Omega = +1$ and
$\Omega = -1$ in the Higgs and confining regimes, respectively. The value of
$\Omega$ thus serves to distinguish the Higgs and confining regimes of the QFTs
we study in this paper.  The fact that the Higgs and confining regimes can be
distinguished in this manner does not necessarily mean that they are separated by a bulk phase
boundary.  While we think the presence of such a phase boundary is plausible, we
have not been able to prove that it must be present.
This will be discussed further in Section~\ref{sec:phase_diagram}.
Fig.~\ref{fig:phase_diag} gives a 
schematic picture of the resulting phase diagram of our theory,
where the red line on which $\Omega$ changes sign may or may not signal a
thermodynamic phase transition.

The fact that vortices may carry a quantized magnetic flux in the theory
we study is similar to the behavior of vortices in charge $N>1$
superconductors, which are gapped phases.
There the parallel result can be understood in terms of topological
order, characterized by a topological $\mathbb{Z}_N$ gauge theory at
long distances.
Here our focus is on a gapless system which does not satisfy any
conventional definition of topological order.  Nevertheless, we explain why it
is natural to view $\Omega$ as a well-defined label for the Higgs and confining
regimes for the class of QFTs we study.
However, as we will discuss, we are currently unable to rule out a scenario where $\Omega$ jumps due to a level-crossing phenomena among multiple types
of vortices in a manner which would not involve a genuine phase transition. Future numerical lattice simulations are needed
to convincingly demonstrate the presence (or otherwise)
of a phase boundary associated with changes in $\Omega$. 

The organization of this paper is as follows.
In Section~\ref{sec:continuum} we
review the continuum model studied in Ref.~\cite{Cherman:2020hbe},
and highlight some limitations and unanswered questions from our
previous work which we now address in Sections \ref{sec:lattice}
and \ref{sec:line_operator_correlators}.
Readers who are happy to accept that $\Omega$ jumps from $-1$ to $+1$ as one
goes from the Higgs to the confining regime can skip ahead to
Section~\ref{sec:phase_diagram} for a discussion of the implications of this
behavior of $\Omega$ for the phase diagram.
Readers who are interested in
detailed evaluations of $\Omega$ can read Section~\ref{sec:lattice},
which introduces a lattice discretization of the model we study
and defines $\Omega$ in terms of line operator correlation functions,
and Section~\ref{sec:line_operator_correlators} which computes $\Omega$
in various limiting regimes.
Section~\ref{sec:conclusions} contains conclusions
and comments on future research directions.
Some ancillary materials are collected in the Appendices.

\section{Higgs versus confinement in a continuum gauge theory } 
\label{sec:continuum} 
 
We consider a $2{+}1$D $U(1)$ gauge theory with monopoles, coupled to
two Higgs scalars $\phi_+$, $\phi_-$ with electric charges $\pm 1$, plus a
neutral complex scalar $\phi_0$. The action is~\cite{Cherman:2020hbe}
\begin{align}
    S = \int d^3x &\bigg[\frac{1}{4g^2} f_{\mu\nu}f^{\mu \nu} +
    |(\partial_{\mu} - i a_{\mu} ) \phi_+|^2 + 
    |(\partial_{\mu} + i a_{\mu} ) \phi_-|^2 + 
    m^2 \left(|\phi_+|^2 + |\phi_-|^2\right) \nonumber \\
    & +  |\partial_{\mu} \phi_0|^2
    + m_0^2 |\phi_0|^2
    +  \epsilon \, (\phi_0 \phi_{+}\phi_{-} + \mathrm{h.c.}) 
    +V_{\rm int}(\phi_+, \phi_-, \phi_0) \bigg] \,.
    \label{eq:continuum_model}
\end{align} 
We assume that the UV completion allows finite-action magnetic monopoles
so that there is no magnetic $U(1)_m$ symmetry.  We also assume that the
potential $V_{\rm int}$ is invariant under the $U(1)$ transformation
$\phi_{\pm} \to e^{i\alpha} \phi_{\pm}$, $\phi_0 \to e^{-2i\alpha} \phi_0
$. The minimal gauge invariant operators such as $\phi_{+} \phi_{-}$ and
$\phi_0^\dagger$ have charge $2$ under this naive $U(1)$ transformation.
The faithfully-acting continuous global symmetry is therefore $U(1)/\ZZ_2$,
under which $\phi_{+}\phi_{-}$ and $\phi_0^\dagger$ have charge $1$.
Finally, we assume that the model enjoys a flavor-flip symmetry
\begin{align}
   \ZF : \begin{cases} 
        a_{\mu} \to - a_{\mu} \\
        \phi_{+} \to \phi_{-} \\
        \phi_{-} \to \phi_{+}
    \end{cases}
\end{align}
so that the internal global symmetry
is
\begin{align}
G =  \ZF \times \left[(U(1)/\ZZ_2) \rtimes (\mathbb{Z}_2)_C \right]\,.
\label{eq:global_sym}
\end{align}
where $(\mathbb{Z}_2)_C$ is the charge conjugation symmetry and acts by $a_{\mu} \to -
a_{\mu}$, $\phi_{\pm} \to \phi_{\pm}^{\dag}$.
In the action (\ref{eq:continuum_model}),
we write explicitly the symmetry-allowed cubic term 
$\phi_0 \, \phi_+ \phi_-$
with coefficient $\epsilon$.
If $\epsilon$ is set to zero
(and $V_{\rm int}$ contains no other terms coupling
$\phi_+\phi_-$ to $\phi_0$), then
the internal global symmetry is enlarged to become
\begin{align}
  G_{\epsilon = 0} =   \left[\frac{U(2)}{U(1)} \times U(1)\right] \rtimes (\mathbb{Z}_2)_C
    = \left[SO(3) \times U(1) \right] \rtimes (\mathbb{Z}_2)_C \,,
\end{align}
where the first factor comes from gauge-inequivalent
rotations acting on the gauge-charged fields, while the second factor comes from
phase rotations of the neutral field.  When $\epsilon \neq 0$, as we will assume
everywhere below, $SO(3) \times U(1)$ is broken to $ \ZF \times (U(1)/\ZZ_2 )$. We will refer to the $U(1)/\ZZ_2$ symmetry as the `baryon number' symmetry 
\begin{align}
\UB \equiv U(1)/\ZZ_2
\end{align}
due to its analogy to the baryon number symmetry in QCD. 

One may interpret the gauge-neutral scalar $\phi_0$ as an interpolating field for
charged-particle bound states like $\phi_{+}^{\dag}\phi_{-}^{\dag}$.  The value
of including $\phi_0$ as an explicit degree of freedom in
Eq.~\eqref{eq:continuum_model} is that it allows us to access a wider range of
regimes of the theory using semiclassical techniques. If not for the inclusion of
$\phi_0$, the model would have only two semiclassically-tractable corners: one
where $m^2/g^4$ is large and negative and the $\UB$ symmetry is spontaneously
broken, and another where $m^2/g^4$ is large and positive and the $\UB$
symmetry is not spontaneously broken.  These are obviously distinct
thermodynamic phases.
But can there be a distinction between Higgs and confining regimes
where $\UB$ is spontaneously broken in \emph{both} regimes?
Including $\phi_0$ in the
story allows us to explore this question by dialing the mass squared parameter
$m_0^2$, or rather the dimensionless combination
$m_0^2/g^4$.   In particular, we can consider parameter regimes where
$m_0^2/g^4$ is held sufficiently negative while $m^2/g^4$ is varied, taking the
theory between confining and Higgs regimes while $\UB$ remains spontaneously broken.
In Ref.~\cite{Cherman:2020hbe} we argued that the physics of global vortices
sharply distinguish these two $\UB$-broken regimes. We showed that in the Higgs
regime unit-winding vortices carry a non-trivial Aharonov-Bohm phase of $\pi$,
and argued that in the confining regime the Aharonov-Bohm phase becomes trivial.   

The argument that the Aharonov-Bohm phase is $\pm\pi$
deep in the Higgs phase,
so that $\Omega = -1$,
is very simple.  The action per unit length of a global vortex
running along some large curve $\Gamma$,
assumed circular for simplicity,
grows logarithmically with the size of the curve and is, for sufficiently
large curves,
dominated by the contributions of scalar kinetic terms.
Provided the magnitude of $\phi_{\pm}$ approaches some constant
$v>0$ far from the vortex, as is the case when $-m^2/g^4 \gg 1$,
and denoting the phases of $\phi_{\pm}$ by $\varphi_{\pm}$,
the Aharonov-Bohm phases of vortices are determined by 
the charged-scalar kinetic terms
\begin{align}
v^2 \left[ (\partial_{\mu} \varphi_{+} 
- a_{\mu})^2+(\partial_{\mu} \varphi_{-} + a_{\mu})^2\right] + \ldots \,.
\end{align}
Thinking of $\Gamma$ as an infinite straight line, and viewing a vortex
$V(\tilde\Gamma)$ as a field configuration where a unit-charge order parameter
winds by $2\pi$ far from $\Gamma$, then $\phi_0$ (which carries charge
$-1$ under $\UB$) must go like $e^{-i \theta}$ far from
the vortex core. Thanks to the cubic $\epsilon$ term in
the action \eqref{eq:continuum_model}, static solutions of the
equations of motion must have one of the phases $\varphi_{\pm}$ approach
$\theta$ far from the vortex,
while the other phase approaches $0$.
And far from $\Gamma$
the gauge field must approach a pure-gauge form,
$a_{\theta} \to \frac{\Phi}{2\pi r}$,
for some value of the enclosed flux $\Phi$.
Minimizing the long-distance
energy density coming from the charged scalar kinetic terms immediately
implies that the magnetic flux carried by the vortex is $\Phi = \pm
\pi$.
\footnote{Our system includes finite-action magnetic
monopoles with magnetic flux $\in 2\pi \mathbb{Z}$, so vortices with $\Phi = \pm
\pi$ mix with each other.  This ensures that the $\ZF$ symmetry is not
spontaneously broken on global vortex worldlines in the
Higgs regime~\cite{Cherman:2020hbe}.} 
Therefore unit-winding vortices carry an Aharonov-Bohm phase factor of
$e^{i\Phi} = \Omega = -1$ deep in the Higgs regime.

In Ref.~\cite{Cherman:2020hbe} we
also argued that the vortex magnetic flux vanishes in the confining
$\UB$-broken regime using a combination of effective field theory ideas
along with physical arguments involving string-breaking effects.
While we believe that these arguments are solid
they are not quantitative, unlike
the calculations we were able to present deep in the Higgs regime,
due to the challenges of
discussing confinement and string-breaking effects using continuum effective
field theory-based techniques.
But taking the result about the
triviality of the Aharonov-Bohm phase in the confining regime at face value, we
obtained a sharp distinction between the confining and Higgs regimes: in the
$\UB$-broken confining regime vortices carry trivial Aharonov-Bohm phases, while
in the Higgs regime they carry non-trivial Aharonov-Bohm phases.

In
Ref.~\cite{Cherman:2020hbe} we argued that the ability to distinguish these two
regimes via a sharp distinction in the physics of vortices provides
an indication of the presence of a genuine phase boundary separating
the Higgs and confining regimes.
However, in that prior work we did not address a number of
related questions.  These include:
\begin{enumerate}
\item
Can this putative order parameter be expressed as a standard
vacuum correlation function?
The quantity studied in Ref.~\cite{Cherman:2020hbe} involved a
Wilson loop expectation value defined by a path integral constrained
to contain a prescribed vortex loop.
Can one extract the physical observable of interest, namely the mutual statistics between electric probes and global vortices, from the correlation function of a Wilson
loop $W(C)$ and some well-defined vortex creation operator $V(\tilde\Gamma)$ localized on $\Gamma$?

\item
Under 
what conditions will \emph{any} long-distance correlation functions
involving the putative vortex-insertion operator $V(\tilde\Gamma)$ be non-trivial?
The issue is that global
vortices on large contours $\Gamma$ of characteristic size $L$
have an action which grows as $\mu \, L \log L$,
with $\mu$ a renormalization point,
or faster than linearly with the loop size.
One should worry that explicit insertions of $V(\tilde\Gamma)$ may get screened by
dynamical vortices.
If so, then
at long distances $V(\tilde\Gamma)$ would flow to the unit operator,
having trivial correlation functions with Wilson loops.
\footnote
    {%
    This is not an issue for the ``local'' vortex line operators in
    superconductors, whose expectations have a perimeter-law fall-off. The perimeter law scaling of both Wilson loops and vortices plays an
    essential role in the robustness of topological order against the addition
    of massive dynamical unit-charge particles, as discussed in, e.g.,
    Ref.~\cite{Cherman:2023xok}. }

\item
Assuming that the issues above are suitably clarified, can one present
a fully explicit and controlled computation demonstrating the
triviality of the Aharonov-Bohm phases of vortices in a confining
$\UB$-broken regime?%
\footnote
    {%
    The behavior of the Aharonov-Bohm flux in
    a similar model was recently examined in
    Ref.~\cite{Hayashi:2023sas},
    which claimed to show that vortices have an Aharonov-Bphm phase of $-1$
    also in the confining phase through an analysis on the
    lattice using a strong coupling expansion.
    However, 
    Ref.~\cite{Hayashi:2023sas} 
    did not sharply define the vortex operator,
    mis-analyzed the confining regime,
    and its conclusion is not reliable.
    }
\end{enumerate}
Addressing these questions will make a much more convincing case for
the existence of a sharp distinction between the Higgs and confining
$\UB$-broken regimes based on the physics of vortices.
Resolving the issues above is also
key for enabling non-perturbative explorations of Higgs-confinement
continuity in $\UB$-broken phases using, e.g., numerical lattice 
simulations.
In this paper we address all of the questions above in the
context of a Euclidean lattice formulation of the continuum model
\eqref{eq:continuum_model},
using a version of the Villain formulation for gauge
and compact scalar fields.
We will see that working on the lattice helps to make the
questions above sufficiently concrete that the answers become very
clear, and provides a natural starting point for future numerical
explorations of the phase diagram using Monte Carlo techniques.
As a start, our analysis leads to the reformulation
\eqref{eq:order_parameter}
of the vortex order parameter discussed in Ref.~\cite{Cherman:2020hbe}.
In Sections
\ref{sec:lattice} and \ref{sec:line_operator_correlators}
we will discuss a lattice construction that allows us to explore
$\Omega$ in various regimes of interest. Then in Section~\ref{sec:phase_diagram}
we explain why our analysis implies that $\Omega$ must change
non-analytically as one varies microscopic parameters 
to get from Higgs to confining regimes within the $\UB$-broken phase.
We also discuss whether such a change in $\Omega$ should be associated with a
bulk phase boundary.

\section{Lattice discretization}
\label{sec:lattice}

\subsection{Vortex operators on the lattice}
\label{sec:vortex_operators}
Since vortex operators play a starring role in our analysis, and the familiar
classification of vortices involves the notion of winding number, we begin this
section by discussing scalar field winding numbers on the lattice. Consider a
continuum field theory of a real compact scalar $\varphi \equiv \varphi +2\pi$.
The winding number $\nu$ on a closed curve $C$ of a field configuration of
$\varphi$ can be written as 
\begin{align}
    \nu(C) = \frac{1}{2\pi}\int_{C} d \varphi \in \mathbb{Z}
    \label{eq:continuum_nu}
\end{align}
To obtain configurations of $\varphi$ for which $\nu(C) \neq 0$ it is crucial to
use the fact that $\varphi$ is compact, and thus needs to be single-valued along
$C$ only up to an integer multiple of $2\pi$.

A common lattice UV completion of a compact scalar involves introducing a
real-valued scalar field on sites $s$, $\varphi_s \in \mathbb{R}$, and
replacing the continuum kinetic term
$\sim \int d^3{x} \> (\partial_{\mu}
\varphi)^2$ by the `XY model' action
\begin{align}
    S_{XY} &= \kappa\sum_{s}\sum_{\mu}  \left[ 
        1 - \cos\left(\varphi_{s+\hat \mu} - \varphi_{s}\right)\right]
	\equiv \kappa\sum_{s}\sum_{\mu}  \left[ 1 -
     \cos\,  \left(d\varphi\right)_{s,\mu} \right],
    \label{eq:XY_action}
\end{align}
where $\kappa$ is the hopping parameter, $s$ labels lattice sites, $\mu$
labels the three orthogonal directions on a cubic lattice with unit lattice
spacing, $\hat\mu$ is a unit vector, and
$(d\varphi)_{s,\mu} \equiv \varphi_{s+\hat\mu} - \varphi_{s}$
denotes a lattice finite difference.
Below we will generally suppresses the individual
$(s,\mu)$ labels for links and refer to them by $\ell$. To reduce clutter we sometimes suppress the lattice labels of fields where it is unlikely to cause confusion. Our lattice
conventions are summarized in Appendix~\ref{sec:lattice_conventions}. 
The naive lattice parallel of Eq.~\eqref{eq:continuum_nu} is 
\begin{align}
    \nu(C) \stackrel{?}{=} \frac{1}{2\pi} \sum_{\ell \in C} \>
    (d\varphi)_{\ell} \,.
    \label{eq:naive_nu}
\end{align}
But this expression is identically zero because $\varphi_s \in
\mathbb{R}$ and hence is single-valued,
implying that $(d\varphi)_\ell$ sums to zero around any closed loop.

If one were to project, by hand, $d\varphi$ onto the range $(-\pi,\pi]$ in
Eq.~\eqref{eq:naive_nu} one could obtain a
non-vanishing result for $\nu(C)$~\cite{Tobochnik:1979zz}.
But this is somewhat arbitrary and unnatural.
For our purposes, it is much nicer to work with a different lattice
discretization in which vortex configurations have a particularly clean definition.
Specifically, we will use a modified form of the Villain lattice action for
scalar
fields~\cite{Villain:1974ir,Gross:1990ub,Sulejmanpasic:2019ytl,Gorantla:2021svj}.
In the Villain formalism, the basic idea is to work with a real-valued field
$\varphi_s \in \RR$ at each site, subject to a $2\pi \ZZ$ gauge redundancy
with an associated discrete gauge field $n_\ell \in \ZZ$. The discrete gauge
transformations take the form
\begin{equation}
     \label{eq:compactness}
\varphi_s \to \varphi_s + 2\pi k_s \,, \quad n_\ell \to n_\ell + (dk)_\ell\,,
\end{equation}
where $k_s \in \ZZ$ is an arbitrary site-dependent gauge parameter.
As a result, the
pair $(\varphi_s, n_{\ell})$ describes a compact scalar field with periodicity
$2\pi$, and one can write a lattice scalar kinetic term as
\begin{align} 
    S_{\textrm{Villain scalar }} =
    \frac{\kappa}{2} \sum_{\ell} \> [ (d\varphi)_\ell - 2\pi n_\ell ]^2  \,,
    \label{eq:villain_scalar}
\end{align}
with the understanding that $\varphi_s$ and $n_\ell$ are independent
fields to be integrated or summed over, respectively.
A path integral based on Eq.~\eqref{eq:villain_scalar} can have a
$U(1)^{(0)}$ 
global symmetry which acts as an overall phase rotation,
$e^{i\varphi_s} \to e^{i\varphi_{s}} e^{i\alpha}$, with
$\alpha \in [0,2\pi)$.
It will be interesting to extend this scalar theory and consider
the closely related model:
\begin{align} 
    S_N = 
    \frac{\kappa}{2} \sum_{\ell} \> [ (d\varphi)_\ell - 2\pi n_\ell ]^2  
    + \frac{1}{2\lambda} \sum_p \> (dn)^2_p 
    - \frac{2\pi i}{N} \sum_p \> v_{\star p} (dn)_p \,.
    \label{eq:villain_scalar_1formZN}
\end{align}
where $\lambda \in \mathbb{R}^{+}$ and $N \in \mathbb{N}$ are new
parameters and $v_{\tilde{\ell}}\in \mathbb{Z}$ is an
auxiliary Lagrange multiplier field
living on links $\tilde \ell$ of the dual lattice.
As discussed further below,
the parameter $\lambda$ controls the size of fluctuations in
the discrete gauge field $n$, while summing over the Lagrange
multiplier $v$ implements a constraint forcing the discrete gauge
flux to vanish modulo $N$ on every plaquette, $(dn)_p = 0 \bmod N$.
With this constraint,
the model has a
$\mathbb{Z}_N^{(1)}$ $1$-form symmetry acting on vortex loop operators
in addition to a $U(1)^{(0)}$ $0$-form symmetry.

In the Villain formulation,
the winding number $\nu(C) \in \mathbb Z$ of the compact scalar
about some contour $C$ is defined as 
\begin{align}
\nu(C)= -\frac{1}{2\pi}\sum_{\ell\in C} 
\left[ (d\varphi)_\ell - 2\pi n_\ell \right]
= \sum_{\ell \in C} n_{\ell} = \sum_{p\in D} (d n)_p \,,
\label{eq:villain_winding}
\end{align}
where the middle equality follows because $\varphi_s$ is single-valued, while
the right-most equality follows from Stokes' theorem assuming that
the contour $C$ is the boundary of some surface $D$.
Unlike the naive definition \eqref{eq:naive_nu},
the winding number
\eqref{eq:villain_winding} may take on any integer value
and will typically be non-zero on generic field
configurations $\{ \varphi_s, n_{\ell} \}$.
Field configurations for which
$\nu(C)$ is non-zero can be interpreted as containing $\nu(C)$ unit-winding
vortices within the contour $C$.

From the above right-most expression for $\nu(C)$ one sees that
the $\lambda$ term in the action \eqref{eq:villain_scalar_1formZN},
involving the square of the field strength of the discrete gauge field
$n_{\ell}$,
controls the per-unit-length action cost of vortex excitations.
The last term in the action \eqref{eq:villain_scalar_1formZN},
involving the Lagrange multiplier $v_{\star p}$,
serves to enforce a constraint eliminating from the path integral
all configurations with winding $\nu(C)\neq 0 \textrm{ mod } N$
about any (topologically trivial) contour.
As a consequence of this constraint, the theory defined by
Eq.~\eqref{eq:villain_scalar_1formZN} has a $\mathbb{Z}_N^{(1)}$ symmetry
generated by the topological line operators
\begin{align}
    U_k(C)  \equiv
    \exp\Bigg(\frac{2\pi i  k}{N}  \sum_{\ell \in C}  n_{\ell} \Bigg) 
    = \exp \Bigg({\frac{2\pi i k }{N} \, \nu(C)}\Bigg),
    \label{eq:one_form_generatorZN}
\end{align}
where $k \in \mathbb Z$ and
$C$ is a closed curve on the lattice. The $2\pi/N $ factor in the exponent
makes $U_k(C)$ topological because $(dn)_p = 0 \textrm{ mod } N$.
One can immediately verify that $U_1(C)^N$ is equal to the identity operator,
and that these operators obey a $\mathbb{Z}_N$ fusion rule,
$U_k(C) \, U_{k'}(C) = U_{k''}(C)$ with $k'' \equiv k + k' \textrm{ mod } N$.
Replacing $2\pi k/N$ by a generic real number in
Eq.~\eqref{eq:one_form_generatorZN} would produce a non-topological line operator.   

We now define vortex insertion operators. The simplest vortex operator is
\begin{align}
    V_{w}(\tilde{\Gamma}) \equiv
    \exp \Bigg(
         \frac{2\pi i w}{N} 
         \sum_{\tilde{\ell}\in \tilde{\Gamma}} \> v_{\tilde{\ell}}  \,
          \Bigg) ,
    \label{eq:vortex_operator}
\end{align}
with $w \in \mathbb Z$ and
$\tilde{\Gamma}$ a closed curve on the dual lattice (for our purposes $\tilde\Gamma$ must be contractible, i.e., the boundary of a 2-surface in the dual lattice).
Inserting $V_w(\tilde{\Gamma})$ into the path integral
shifts the value of the constraint enforced by the Lagrange
multiplier field $v$;
a short computation shows that 
\begin{equation}
    \nu(C) \bmod N
    =  w \, \textrm{Link}(C,\tilde{\Gamma}) \textrm{ mod } N \,.
    \label{eq:vorticity1}
\end{equation}  
This justifies the
interpretation of Eq.~\eqref{eq:vortex_operator} as a vortex insertion
operator, and implies that $V_w(\tilde{\Gamma})$ carries charge
$w \textrm{ mod } N$ under the
$\mathbb{Z}_N^{[1]}$ symmetry, so that
\begin{align}
    \left\langle U_k(C) \, V_w(\tilde{\Gamma}) \right\rangle
    = e^{\frac{2\pi i kw}{N} \textrm{Link}(C, \tilde{\Gamma})} \,
    \langle U_k(C) \rangle \,
    \langle V_w(\tilde{\Gamma}) \rangle \,.
    \label{eq:ZN1_action}
\end{align}

As reviewed in Appendix~\ref{sec:AH_appendix},
particle-vortex duality holds exactly in this Villain formulation
lattice theory.
This duality maps the above theory to a charge-$N$ Abelian-Higgs model
without monopoles, with scalar hopping parameter $\lambda/N^2$ and the
squared gauge coupling $(2\pi)^2\kappa$.
The global vortex operators \eqref{eq:vortex_operator} (with slight modifications when $\lambda$ is finite) map to Wilson lines. The absence of configurations with $\nu(C)\neq 0 \bmod N$ in the
scalar field theory \eqref{eq:villain_scalar_1formZN} corresponds
to the absence of excitations with electric charge $q \neq 0 \bmod N$
in the dual charge-$N$ Abelian-Higgs model.

We now consider promoting the integer-valued Lagrange multiplier $v_{\tilde\ell}$ to a real-valued $\theta_{\tilde\ell} \in \RR$, 
\begin{align} 
    S = 
    \frac{\kappa}{2} \sum_{\ell} \> [ (d\varphi)_\ell - 2\pi n_\ell ]^2  
    - i\sum_{\tilde{\ell}} \theta_{\tilde{\ell}} \> (dn)_{\star \tilde{\ell}}  \,,
    \label{eq:villain_scalar_1form}
\end{align}
This action can be thought of as describing the $N \to \infty$ limit of $S_N$ \eqref{eq:villain_scalar_1formZN}.
This model now has a continuous $U(1)^{(1)}$ symmetry
generated by topological line operators
\begin{align}
    U_\alpha(C)  \equiv
    \exp\bigg(i \alpha \sum_{\ell \in C}  n_{\ell} \, \bigg)  ,
    \label{eq:one_form_generator}
\end{align}
with $\alpha \in \mathbb R$.
This symmetry
acts on vortex operators with winding number $w \in \mathbb Z$,,
\begin{align}
    V_w(\tilde{\Gamma})
    = \exp\bigg(i w \sum_{\tilde{\ell} \in \tilde{\Gamma}}\theta_{\tilde{\ell}} \bigg) ,
\end{align}
in the same manner as \eqref{eq:ZN1_action}, namely
\begin{align}
    \left\langle U_\alpha(C) \, V_w(\tilde{\Gamma}) \right\rangle
    = e^{i \alpha w \> \textrm{Link}(C, \tilde{\Gamma})} \,
    \langle U_\alpha(C) \rangle \,
    \langle V_w(\tilde{\Gamma}) \rangle \,.
    \label{eq:Z1_action}
\end{align}
(If $C$ is contractible then
$
    \langle U_\alpha(C) \rangle
$
equals unity.)

The $N \to \infty$ theory \eqref{eq:villain_scalar_1form}
is dual to a pure $U(1)$ gauge
theory without monopoles, and has a $U(1)^{(1)}$ $1$-form symmetry
associated with conservation of magnetic flux
in
addition to a $U(1)^{(0)}$ symmetry.  These two symmetries have a mixed
't Hooft anomaly \cite{Gaiotto:2014kfa,Delacretaz:2019brr},
as discussed in Appendix \ref{sec:anomaly_appendix}.
The generalized Coleman-Mermin-Wagner
theorem for $1$-form symmetries
\cite{Gaiotto:2014kfa,Lake:2018dqm} says that $U(1)^{(1)}$ symmetries cannot
be spontaneously broken in $2{+}1$ dimensions,
and thus the 't Hooft anomaly is matched by the spontaneous breaking of the
$U(1)^{(0)}$ shift symmetry of \eqref{eq:villain_scalar_1form} for all $\kappa$
\cite{Gaiotto:2014kfa}.

As an aside, it is instructive to consider the special case of
of the theory \eqref{eq:villain_scalar_1formZN} when
$N = 1$.
Then there is no imposed constraint on $(dn)_p$ and
field configurations with all winding numbers
contribute to the path integral.
Our putative vortex insertion operator $V_{w}(\tilde{\Gamma})$
defined in \eqref{eq:vortex_operator}
becomes the trivial identity operator.
Nevertheless,
one may attempt to define a non-trivial vortex operator
in a different manner
by modifying the $(dn)_p^2$ term in the action,
\begin{align}
\frac{1}{2\lambda} \sum_p (dn)_p^2  
\to \frac{1}{2\lambda} \sum_p \left[(dn)_p - w \,[\tilde \Gamma]_{\star p}\right]^2\,,
\label{eq:other_vortex_operator} 
\end{align}
so as to preferentially bias the action toward configurations having
non-vanishing discrete gauge flux $w$ on plaquettes dual to links in
$\tilde \Gamma$.%
\footnote
    {%
    Here and henceforth, $[\tilde\Gamma]_{\star p}$ denotes the
    characteristic function of $\tilde\Gamma$
    equal to 1 for $\star p \in \tilde\Gamma$, 0 otherwise.
    }
In fact, this is precisely the particle-vortex dual of a charge-$w$
electric Wilson loop in the charge-$1$ Abelian-Higgs model,
see Appendix~\ref{sec:AH_appendix}. The
above operator creates a per-unit-length incentive for the system to create
vorticity around the curve $\tilde\Gamma$.
Although this change
\eqref{eq:other_vortex_operator} is non-trivial,
viewed as an operator insertion it 
flows to the identity operator at long distances.
To see how this works, note
that the leading cost in action of a dynamical vortex of size $L$ grows
like $L \log L$.
When $N=1$, for large enough $L$ (with fixed $\lambda$ and $\kappa$)
the $L\log L$ cost of a vortex outweighs the perimeter-law incentive for
vortex creation produced by the change
\eqref{eq:other_vortex_operator}.
Consequently,
for sufficiently large $L$, the bias in the action produced
by the vortex creation operator \eqref{eq:other_vortex_operator} will 
be insufficient to actually create a vortex.
Equivalently but more descriptively,
the system will produce a dynamical
anti-vortex that screens the vortex we inserted by hand.
This results in a vanishing winding number as measured on
large contours $C$ which link with $\tilde{\Gamma}$.
And hence, when $N=1$ and there is no $1$-form symmetry, all vortex
operators have trivial correlation functions in the long-distance limit.

The lesson we take from this discussion is that if we wish to
formulate an order parameter which involves \emph{long-distance}
correlation functions of vortex line operators,
we should focus on doing so in theories formulated to have a
$1$-form symmetry which protect vortices from being completely screened,
as is the case in
our theory \eqref{eq:villain_scalar_1formZN} with $N > 1$.
And the simplest choice is to send $N \to \infty$, producing
the theory \eqref{eq:villain_scalar_1form} in which no vortex
screening whatsoever can take place.

Therefore, in the following sections
we will ensure that the more complicated theories we study
have a $U(1)^{(1)}$ $1$-form symmetry, with one sector
looking just like \eqref{eq:villain_scalar_1form}.  As we already
mentioned, continuous $U(1)^{(1)}$ symmetries cannot be spontaneously
broken in $2{+}1$D, so this $U(1)^{(1)}$ will be a spectator in the
dynamics we discuss below, but its presence will allow us to have the
cleanest possible definition of the observables we are interested in.
This addresses point 2 raised in Sec.~\ref{sec:continuum}.

\subsection{Coupled scalar-Higgs theory}
\label{sec:model_and_line_operators}

Keeping in mind the preceding discussion,
we now construct a lattice field theory analog of
the continuum theory
\eqref{eq:continuum_model}.
We write our lattice action as
\begin{align}
    S = S_{\rm QED} + S_{\rm neutral} + S_{\rm mixing} \,,
\end{align}
where $S_{\rm QED}$ is a Villain lattice action for two-flavor scalar QED,
$S_{\rm neutral}$ is a Villain lattice action for an electrically-neutral scalar
field, and $S_{\rm mixing}$ couples the neutral scalar and the charged scalars
in such a way that there is a single continuous global symmetry $\UB$. For an
appropriate choice of microscopic parameters below there will also be also $\ZF$
charged-scalar-exchange symmetry. We assume that the full action $S$ also has a
standard coordinate reflection symmetry, which forbids a Chern-Simons term for
the gauge field $a$, as well as a charge conjugation symmetry which will act by
flipping the sign of all of the dynamical fields. Finally, there will be a
$U(1)^{(1)}$ symmetry that acts on vortex line operators, so that vortex
correlation functions with Wilson loops can be non-trivial in the long-distance
limit.

The QED action $S_{\rm QED}$ takes the form
\begin{align}
    S_{\rm QED} \equiv &\, \frac{\beta}{2} \sum_{p} \>
    [(da)_{p} - 2\pi m_{p}]^2  
    +   \sum_{\pm }\frac{\kappa_{\pm}}{2} \sum_{\ell} \>
    [ (d \varphi_{\pm})_{\ell} \mp \,  a_{\ell} - 2\pi  (n_{\pm})_{\ell} ]^2\,.
\label{eq:action_a_chi_pm}
\end{align}
The pair of fields $a_\ell \in \mathbb{R}$ and $ m_p \in \mathbb{Z}$ describe a
compact $U(1)$ gauge field with gauge coupling $e^2 \equiv 1/\beta$.
The magnetic flux through any closed 2-surface $M_2$ is quantized,
\begin{equation}
    \mathrm{flux}(M_2) \equiv
    \sum_{p \in M_2} \left[(da)_p - 2\pi m_p\right]
    \in 2\pi \mathbb{Z} \,.
\end{equation}
Similarly,
the fields $(\varphi_{\pm})_s \in \RR$ and
$(n_{\pm})_\ell \in \ZZ$ describe a pair of
charge $\pm 1$ compact scalar fields
having hopping parameters $\kappa_{\pm} \in \mathbb{R}_{+}$.
Setting $\kappa_{+} = \kappa_{-}$ endows the theory
with an additional $\ZF$ symmetry (analogous to the flavor flip symmetry discussed in section \ref{sec:continuum}) that 
interchanges the charged scalars and negates the gauge field,
\begin{align}
    \ZF:  \;
    (\varphi_{\pm})_s \to (\varphi_{\mp})_s, \quad
    (n_{\pm})_{\ell} \to  (n_{\mp})_{\ell}, \quad
    a_{\ell} \to - a_{\ell}, \quad
    m_p \to -m_p  \,.
\end{align}
The gauge redundancies associated with $S_{\rm QED}$ are
\begin{subequations}
\label{eq:higgsgaugetransform}
\begin{align} 
    a_\ell &\to a_\ell + (d\alpha)_\ell + 2\pi r_\ell,
    &m_p &\to m_p + (dr)_p,
\\
    (\varphi_{\pm})_s &\to (\varphi_{\pm})_s \pm \alpha_s + 2\pi (k_\pm)_s,
    &(n_\pm)_\ell &\to (n_\pm)_\ell + (dk_\pm)_\ell \mp r_\ell,
\end{align} 
\end{subequations}
where $\alpha_s \in \RR$, $r_\ell \in \ZZ$ and $(k_\pm)_s \in \ZZ$
are arbitrary gauge parameters.

There is no magnetic $0$-form symmetry in $S_{\rm QED}$. Magnetic
monopole-instanton configurations are associated with configurations with
non-vanishing discrete flux through cubes,
$(d m)_c \neq 0$.
As we sum over all values of $m_{p}$,
monopole-instanton configurations have finite action.
(One could control the action of monopole configurations
by dialing the coefficient of an additional interaction term
in the action of the form $\sum_c (dm)_c^2$.
To keep things simple, we do not add such a term.)

Our action for the neutral scalar field is
\begin{align}
    S_{\rm neutral} \equiv
    \frac{\kappa_0}{2} \sum_{\ell} \>
    [ (d\varphi_0)_\ell - 2\pi (n_0)_\ell ]^2  
    - i\sum_{\tilde{\ell}} \>
    \theta_{\tilde{\ell}} \, (dn_0)_{\star \tilde{\ell}} \,,
\end{align}
where the lattice variables $(\varphi_0)_s \in \mathbb{R}$,
$(n_0)_{\ell} \in \mathbb{Z}$, and
$\theta_{\tilde{\ell}} \in \mathbb{R}$ have
the gauge redundancies:
\begin{equation}
(\varphi_0)_s \to (\varphi_0)_s + 2\pi (k_0)_s 
\,, \quad 
(n_0)_\ell \to (n_0)_\ell + (dk_0)_\ell
\,, \quad
\theta_{\tilde{\ell}} \to \theta_{\tilde{\ell}} + (d\sigma)_{\tilde\ell} + 2\pi h_{\tilde{\ell}} \,,
\end{equation}
where $(k_0)_s,h_{\tilde\ell} \in \ZZ$ and $\sigma_{\tilde s}\in \RR$. This model has a $U(1)^{(1)}$ symmetry
generated by topological line operators
of the form \eqref{eq:one_form_generator} (with $n \to n_0$).
Once again, this symmetry
acts on winding $w$ vortex operators,
\begin{align}
    V_w(\tilde{\Gamma})
    = \exp\bigg(i w \sum_{\tilde{\ell} \in \tilde{\Gamma}}\theta_{\tilde{\ell}}\bigg) \,,
\end{align}
in the manner shown in Eq.~\eqref{eq:Z1_action}.
To simplify notation, subsequently we will often
write just $V(\tilde\Gamma)$ when discussing minimal
winding vortices with $w = 1$.
In the rest of the paper we will analyze the correlation functions of these
vortex operators with unit-charge Wilson loops given by
\begin{align}
    W(C) = \exp\bigg(i \sum_{\ell \in C} a_{\ell} \bigg)\,.
\end{align}

Finally, the $S_{\rm mixing}$ term in the action couples
the neutral scalar $\varphi_0$ to the charged scalars
$\varphi_{+}$ and $\varphi_{-}$, via
\begin{align} 
    S_{\rm mixing} \equiv \frac{\epsilon}{2} \sum_s
    \left[(\varphi_{0})_s - (\varphi_{+})_s - (\varphi_{-})_s - 2\pi t_s
    \right]^2 \,,
\end{align}
where $\epsilon >0$ is a real parameter.%
\footnote
    {%
    Setting $\epsilon$ to zero would cause the theory to acquire
    an unwanted additional continuous $0$-form global symmetry
    associated with independent shifts of $\varphi_{0}$ and
    $\varphi_{+} + \varphi_{-}$.
    Our interest lies in the theory with $\epsilon > 0$.
    }
The newly introduced lattice variable  $t_s \in \ZZ$ is a
discrete dynamical field whose role is to
maintain the $2\pi$ periodicity of the compact scalar fields.
Accordingly, 
\begin{equation}
    t_s \to t_s + (k_0)_s  -  (k_{+})_s - (k_{-})_s
\end{equation}
under the gauge transformations 
\eqref{eq:compactness} and \eqref{eq:higgsgaugetransform}.

This form of $S_{\rm mixing}$ is chosen with two goals in mind.
First, just like the $\epsilon$ term in the continuum model
\eqref{eq:continuum_model},
$S_{\rm mixing}$ ensures that there is only one continuous global symmetry,
which acts as 
\begin{align}
    \varphi_{\pm} \to \varphi_{\pm} + \alpha\,, \quad
    \varphi_0 \to \varphi_0 - 2\alpha \,,
\end{align}
where the gauge-inequivalent choices of $\alpha \in [0,\pi)$ parameterize
$\UB \equiv (U(1)/\ZZ_2)^{(0)}$.
As a result, excitations created by $e^{i\varphi_{0}}$ will mix
with bound states of $e^{i\varphi_{+}}$ and $e^{i\varphi_{-}}$ excitations.
The second goal ensured by this particular form of $S_{\rm mixing}$
is that this mixing takes place through a quadratic term,
which simplifies our later analysis.
As discussed above in Sec.~\ref{sec:vortex_operators},
the mixed 't Hooft anomaly between the
$U(1)^{(1)}$ and $\UB$ symmetries implies that
the $\UB$ symmetry is necessarily broken throughout
the parameter domain where
$\kappa_0$, $\kappa_{\pm}$, $\beta$, and $\epsilon$
are all non-zero and positive.

For later reference, we write a self-contained single expression for
the complete lattice action $S$:
\begin{align}  
    S &=\frac{\kappa_0}{2} \sum_{\ell} \>
	[ (d\varphi_{0})_\ell - 2\pi (n_0)_\ell ]^2 
     +  i\sum_{p} \theta_{\star p} \, (dn_0)_{p}
     + \frac{\beta}{2} \sum_{p} \, [(da)_{p} - 2\pi m_{p}]^2
 \nonumber \\
      &+  
       \sum_{\pm } \frac{\kappa_{\pm}}{2}
       \sum_{\ell} \,
       [ (d \varphi_{\pm})_{\ell} \mp a_{\ell} - 2\pi  (n_{\pm})_{\ell} ]^2
       +  \frac{\epsilon}{2} \sum_s 
    \left[(\varphi_{0})_s - (\varphi_{+})_s - (\varphi_{-})_s - 2\pi t_s
    \right]^2 ,
\label{eq:full_action}
\end{align}
with $(\varphi_0)_s$, $(\varphi_\pm)_s$, $a_\ell$,
$\theta_{\tilde\ell} \in \mathbb R$
and $(n_0)_\ell$, $(n_\pm)_\ell$, $m_p$, $t_s \in\mathbb Z$.
The full gauge redundancies are:
\begin{subequations}
\label{eq:full_gauge_transformations}
\begin{align} 
    (\varphi_{0})_s &\to (\varphi_{0})_s + 2\pi (k_0)_s \,,
    &(n_0)_\ell &\to (n_0)_\ell + (dk_0)_\ell\,,
\\
    (\varphi_{\pm})_s &\to (\varphi_{\pm})_s \pm \alpha_s + 2\pi (k_\pm)_s\,,
    &(n_\pm)_\ell &\to (n_\pm)_\ell + (dk_\pm)_\ell \mp r_\ell\,,
\\
    a_\ell &\to a_\ell + (d\alpha)_\ell + 2\pi r_\ell\,,
    &m_p &\to m_p + (dr)_p,
\\
    t_s &\to t_s + (k_0)_s  -  (k_{+})_s - (k_{-})_s\,,
    &\theta_{\tilde\ell} &\to \theta_{\tilde\ell}+(d\sigma)_{\tilde\ell} + 2\pi h_{\tilde\ell} \,,
\end{align}
\end{subequations}
with $(k_0)_s$, $(k_\pm)_s$, $r_\ell$, $h_{\tilde\ell} \in \mathbb Z$
and $\alpha_s, \sigma_{\tilde s} \in \mathbb R$.

\section{Line operator correlation functions}
\label{sec:line_operator_correlators}

In this section we compute the correlation functions of the line
operators we use to examine the phase structure of our model.
We first discuss symmetry constraints on relevant correlators 
in Sec.~\ref{sec:sym_constraints}
and then derive a useful dual representation of our model
in Sec.~\ref{sec:dual_representation}.
Then we compute the line correlation operator correlation functions,
first in the $\UB$-breaking confining regime in
Sec.~\ref{sec:U1_breaking_confining_regime}, and then in the
$\UB$-breaking Higgs regime in Sec.~\ref{sec:U1_breaking_Higgs_regime}.

\subsection{Symmetry implications}
\label{sec:sym_constraints}

To begin, it will be helpful to consider the constraints on
correlators arising from Hermiticity as well as
the charge conjugation and $\ZF$ symmetries.
Because of the presence of
the imaginary $i\,\theta_{\star p} (dn_0)_p$ term in
the action \eqref{eq:full_action},
the usual
Hermiticity statement that
$
    \langle \mathcal O \rangle^*
    =
    \langle \mathcal O^* \rangle
$
is only valid for
observables $\mathcal O$ which are independent
of the Lagrange multiplier $\theta_{\tilde\ell}$.
More generally,
$
    \langle \mathcal O(\theta) \rangle^*
    =
    \langle \mathcal O(-\theta)^* \rangle
$.
This implies that vortex loop expectations are real,
\begin{align}
    \langle V_w(\tilde{\Gamma}) \rangle \in \mathbb{R} \,,
\end{align}
and that
\begin{align}
    \langle W(C) \, V_w(\tilde{\Gamma}) \rangle^*
    =
    \langle W(C)^* \, V_w(\tilde{\Gamma}) \rangle  \,.
\label{eq:WV*}
\end{align}
Note that 
$
    \langle V_w(\tilde{\Gamma}) \rangle
$
is also necessarily positive since,
after integrating over $\theta_{\star p}$,
the resulting measure is real and positive.

Unbroken charge conjugation symmetry%
\footnote
    {%
    Because the $\UB$ symmetry is spontaneously broken,
    the charge conjugation transformation which leaves
    a particular choice of spontaneously broken vacuum
    invariant is
    the naive charge conjugation transformation
    (which flips the sign of all our lattice fields)
    conjugated by a $U(1)$ phase rotation dependent
    on the particular choice of vacuum.
    This extra conjugation is not relevant
    for vortex or Wilson loop operators.
    }
implies that
vortex expectations are independent of the direction of winding
and that Wilson loop expectations are real,
\begin{align}
    \langle V_w(\tilde{\Gamma}) \rangle
    =
    \langle V_{-w}(\tilde{\Gamma}) \rangle \,,\quad
    \langle W(C) \rangle \in \mathbb{R} \,.
\end{align}
The $\ZF$ flavor-flip symmetry also converts
$W(C)$ to $W(C)^*$ while leaving $V_w(\tilde\Gamma)$
unchanged.
Hence this symmetry, if not spontaneously broken,
also shows that $\langle W(C) \rangle$ is real and,
when combined with \eqref{eq:WV*}, implies that
\begin{align}
    \langle  W(C) \, V_w(\tilde{\Gamma})\rangle \in \mathbb{R} \,.
\end{align}
Consequently, the phase of the $w = 1$ correlator,
\begin{align}
    \Omega \equiv
    \frac{\langle  W(C) \, V(\tilde{\Gamma})\rangle}
    {|\langle  W(C) \, V(\tilde{\Gamma})\rangle|} \,,
\end{align}
which defines the Aharonov-Bohm flux of minimal winding vortices,
can only be $\pm1$,
showing that
the magnetic flux carried by vortices must be either
$\pi \bmod 2\pi$, or $0 \bmod 2\pi$
(provided the $\ZF$ symmetry is unbroken).

It will be easy to see that the
$\ZF$ symmetry of our theory is not spontaneously broken in
any of the semi-classically tractable regions of parameter space,
so we will find that $\Omega = \pm 1$ in our analysis
below of controllable parameter regions.
If $\ZF$ were to break spontaneously in some strongly-coupled
interior region, as sketched in Fig.~\ref{fig:phase_diag},
this region would necessarily be accompanied by a thermodynamic
phase transition, and so would not cause any issue with our goals.

In the remainder of this  section
we will focus our attention on the physics of our theory
in the limit of $\kappa_0 \gg 1$ while considering various
scalings for $\kappa_{\pm}$ and $\beta$.
Holding $\kappa_0 \gg 1$ simplifies our calculations
(and also allows us to take a continuum limit if we scale 
$\kappa_{\pm}$, $\beta$ appropriately),
but due to the fact that $\UB$ is spontaneously
broken for all choices of parameters in our model,
decreasing $\kappa_0$ to $O(1)$ values is not expected
to lead to any significant changes.

\subsection{Worldvolume representation}
\label{sec:dual_representation}

It will be useful to introduce 
dual representations of our model,
which we will refer to as ``worldvolume representations,''
in which the gauge field is ``integrated out'' and replaced
by a sum over appropriate worldsheets and (some of) the matter
fields are similarly replaced by sums over worldlines.
Such representations in Abelian lattice gauge theories 
have a long history; see e.g., Ref.~\cite{RevModPhys.52.453} for a review.
Dualizing the gauge field results in a representation 
\begin{align}
    S = S_{\rm neutral} + S_{\rm worldsheet} + S_{\rm mixing}
    \label{eq:worldsheet_full_action}
\end{align}
where $S_{\rm neutral}$ and $S_{\rm mixing}$ have the same form as in the
preceding section, while the gauge theory action $S_{\rm QED}$ is replaced by
\begin{align} \label{eq:gauge_worldsheet}
    S_{\rm worldsheet}(\Sigma,\varphi_c,n_c)
    = \frac{1}{2\beta} \sum_p \> [\Sigma]_p^2
    &+ \frac{\kappa_c}{2}\sum_\ell \> [(d\varphi_c)_\ell-2\pi (n_c)_\ell]^2
    + \frac{\kappa_c}{2\kappa_+ \kappa_-}\sum_\ell \> [\partial\Sigma]_\ell^2
\nonumber\\ &
    +2\pi i \, \frac{\kappa_c}{\kappa_-}\sum_\ell \> (n_c)_\ell \, [\partial\Sigma]_\ell  \,,
\end{align}
We relegate the explicit derivation of Eq.~\eqref{eq:gauge_worldsheet} to
Appendix~\ref{sec:worldline_action}.
Here $\Sigma$ represents a worldsheet, that is
an arbitrary set of plaquettes which may be viewed as forming one or more
two-dimensional electric-flux surfaces (open or closed).
The path integral now involves a sum over all possible collections
of open and closed surfaces $\Sigma$.
The worldsheet boundaries $\partial\Sigma$ represent worldlines
of the underlying charged particles
(generated by a hopping parameter expansion in $\varphi_+ - \varphi_-$).
The fields $\varphi_c \in \RR$ and $n_c \in \ZZ$ appearing above represent
the remaining electrically-neutral composite scalar field, with
$\varphi_c \equiv \varphi_{+}+\varphi_{-}$ and
$n_c \equiv n_+ + n_-$.
The composite hopping parameter $\kappa_c$ is given by
\begin{align}
    \kappa_c \equiv \frac{\kappa_+ \kappa_-}{\kappa_++\kappa_-}\,.
\end{align}

The theory defined by the action \eqref{eq:gauge_worldsheet} is
exactly dual to our prior theory \eqref{eq:full_action} for any choice of parameters,
but it is especially useful in the lattice strong-coupling limit, $\beta \ll 1$,
where large worldsheets are highly suppressed.
So long as $\beta \ll 1$, this representation provides a useful
starting point for an analysis of the physics in both regimes
$\kappa_{\pm} \ll 1$ and $\kappa_{\pm} \gg 1$.

\subsection{$\UB$-breaking confining regime}
\label{sec:U1_breaking_confining_regime}

Consider the parameter regime $\kappa_{\pm} \ll 1$ and $\beta \ll 1$.
Taking $\kappa_{\pm} \ll 1$ means that the
$\varphi_{\pm}$ particles are heavy in lattice units,
while $\beta \ll 1$ means the lattice gauge dynamics is strongly coupled
(i.e., far from continuum perturbative behavior).
In this regime, charged test particles are confined in the same
heuristic sense as probe quarks are confined in QCD.
We emphasize that taking $\kappa_{\pm}, \beta \ll 1$ means
that in this section we are examining physics far from the continuum limit.

Given that $\kappa_\pm \ll 1$ implies that $\kappa_c$ is also tiny,
it will be useful to work with a further dual
representation of the model in which the composite fields $\varphi_c$, $n_c$
are also dualized in terms of neutral worldlines.
This will produce a representation in which, in addition to the
electric flux worldsheet $\Sigma$ one also sums over
a collection $\Xi$ of worldlines of the neutral composite scalar $\varphi_c$.
For details, see Eq.~\eqref{eq:worldvolume_action}.

It will also be helpful for our following discussion
to incorporate in the effective action 
the effect of inserting a Wilson loop $W(C)$ and a vortex loop $V(\tilde{\Gamma})$
into the original path integral.
The discussion in Appendix~\ref{sec:worldline_action} shows that inserting a contractible
charge-$q$ Wilson loop $W_q(C)$ into the path integral corresponds to a change
of the dual action
\begin{align}
    \frac{1}{2\beta} \sum_p  \>
    [\Sigma]_p^2
    \ \to \ 
    \frac{1}{2\beta}  \sum_p \> ([\Sigma]_p- q[D]_p)^2\,,
\end{align}  
where $D$ is some surface which spans the loop $C$, so
$\partial D = C$.
The resulting worldvolume action is
\begin{align}  
    \label{eq:confining_eff_action}
    S_{\rm eff}(\Sigma,\Xi, \varphi_0, n_0; C, \tilde{\Gamma}) 
    &=
    \frac{\kappa_0}{2}\sum_\ell \left[ (d\varphi_0)_\ell - 2\pi (n_0)_\ell \right]^2
    + i\sum_p \theta_{\star p} \left[ (dn_0)_p - (\star[\tilde \Gamma])_p \right]
\nonumber\\
    &+ \frac{1}{2\beta} \sum_p \> ([\Sigma]_p-[D]_p)^2 
    + \frac{1}{2\kappa_c}\sum_\ell
	\Bigl([\Xi]_\ell - \frac{\kappa_c}{\kappa_-} \, [\partial\Sigma]_\ell \Bigr)^2
\nonumber \\
    &+ \frac{\kappa_c}{2\kappa_+ \kappa_-}\sum_\ell \> [\partial\Sigma]_\ell^2  
    + \sum_s \left[\frac{1}{2\epsilon} \, [\partial \Xi ]_s^2
	+ i (\varphi_0)_s \, [\partial \Xi]_s\right] ,
\end{align}
where we have also included
a minimal winding vortex loop insertion $V(\tilde{\Gamma})$.
In this representation, one sums over surfaces $\Sigma$ and curves $\Xi$,
where $\Sigma$ is a collection of worldsheets of electric flux whose boundaries
$\partial\Sigma$ represent charged particle worldlines, and $\Xi$ is a collection
of worldlines of the electrically-neutral composite particles built out of
$\varphi_+{+}\varphi_-$.
The last term in the effective action indicates that the
endpoints of composite worldlines $\partial \Xi$ are charged under $\UB$
and must be dressed by insertions of $e^{i\varphi_0}$.%
\footnote
    {%
    Provided $\epsilon \ll {|s-s'|}/(2\kappa_c)$, the
    two-point function of the composite field is given by $\langle
    e^{i(\varphi_c)_s}e^{-i(\varphi_c)_{s'}} \rangle =
    e^{-\frac{1}{\epsilon}}\langle e^{i(\varphi_0)_s}e^{-i(\varphi_0)_{s'}}
    \rangle$ to leading order in $\epsilon$ and $\kappa_c$. This indicates
    that even in the confining regime the heavy composite field `tracks'
    the neutral field due to the $\epsilon$ term. This does not, however,
    imply that the integer-valued Villain fields $n_c$ and $n_0$ are tightly
    correlated.
    }

Given the representation \eqref{eq:confining_eff_action}, it is straightforward
to deduce the behavior of the expectations
$\langle W(C)\rangle$,
$\langle V(\tilde{\Gamma})\rangle$ and
$\langle W(C) V(\tilde{\Gamma})\rangle$.
Let us start with $\langle W(C)\rangle$
and, for simplicity, consider the $\ZF$ symmetric case 
$\kappa_{\pm} = 2 \kappa_c$.
Since worldsheet boundaries $\partial\Sigma$ are highly suppressed,
one contribution to $\langle W(C)\rangle$ will arise from the term
where $\Sigma$ is closed and $\Sigma - D$ is a minimal spanning surface
of the contour $C$, and $\Xi$ is empty.
This contribution gives
\begin{align}
    \langle W(C)
\rangle \sim e^{-\frac{1}{2\beta} A[C]} \,, 
\end{align} 
where $A[C]$ is the minimal spanning area of $C$.
This confining area-law behavior
will be the dominant contribution for sufficiently small loops $C$
(or arbitrarily large but fixed loops in the limit $\kappa_\pm \to 0$),
where string-breaking contributions from the matter fields are
negligible due to the small values of $\kappa_{\pm} \ll 1$,
reflecting the fact the electrically-charged fields are very heavy.

One may instead consider a very large contour $C$ when $\kappa_\pm$ is
small but fixed.
For sufficiently large Wilson loops
it becomes advantageous to allow
$\Sigma$ to have boundaries.
Indeed for asymptotically large contours it is
advantageous to minimize the $\beta$ term in the action by setting $\Sigma = D$,
so that $\partial\Sigma = C$.  Therefore in this limit the charged matter
worldlines screen the Wilson loop.  The effective action evaluated on this
leading large-loop screening configuration takes the form
\begin{align} 
    S_{\rm eff}(\Sigma {=} D, \Xi,\varphi_0,n_0; C) &= \frac{\kappa_0}{2}\sum_\ell  [(d\varphi_{0})_\ell - 2\pi (n_0)_\ell]^2 + i
\sum_p   \theta_{\star p} (dn_0)_p
        \label{eq:S_eff_W}
\\ &+
    \frac 1{2\kappa_c}
    \sum_{\ell} \left[
     \tfrac{1}{4} [C]_\ell^2 
    +\left([\Xi]_\ell - \tfrac{1}{2}[C]_\ell\right)^2 \, \right] + 
   \sum_s \left[\frac{1}{2\epsilon}[\partial \Xi]_s^2 
   + i (\varphi_{0})_s[\partial \Xi]_s  \right] . 
\nonumber
\end{align}
Taking $\Sigma$ to coincide with $D$ means
that the would-be confining string is completely broken, and one expects perimeter-law
behavior for $\langle W(C)\rangle$.
To see this, and determine the value of the screening mass $m$ which
will multiply the perimeter $L(C)$ of the Wilson loop contour $C$,
requires understanding which composite worldlines $\Xi$ will be dominant. 
To do so, we first
consider contributions to the path integral where the
composite worldlines $\Xi$ are closed.
Then there are two potential dominant
contributions: one where $\Xi = 0$ and another where $\Xi = C$.
For both configurations $\partial \Xi = 0$ so
the last terms in \eqref{eq:S_eff_W} do not contribute.

These two configurations give identical perimeter-law contributions,
so that
\begin{align}
    \langle W(C)\rangle = 2 \, e^{-m L[C]} + \cdots \,,
    \label{eq:W_leading}
\end{align}
with an leading order screening mass $m = (4\kappa_c)^{-1}$
and the ellipsis ($\cdots$) denoting further contributions from other
worldlines $\Xi$ and worldsheets $\Sigma$.
These further contributions 
fall into two basic classes:
\begin{enumerate} 
    \item There are subleading contributions involving fluctuations of
        $\Sigma$ surfaces and closed $\Xi$ worldlines, which lead to
        corrections to the screening mass $m$
	and the pre-exponential factor of \eqref{eq:W_leading}
	suppressed by factors of
        $e^{-1/2\beta}$ and $e^{-1/2\kappa_c}$, respectively. 
    \item There are corrections from open $\Xi$ worldlines, whose endpoints
        are sources for the neutral scalar $\varphi_0$. The leading
	contributions of this type are single line segments stretched
	along $C$.  These contributions carry no additional
	$\kappa_c$-suppression, but each such line segment
        comes with a suppression factor of $e^{-1/\epsilon}$. 
\end{enumerate}
The sum of such corrections from a single open worldline
segment connecting sites $s,\,s' \in C$ gives
\begin{align}
    \langle W(C)\rangle = 
    e^{-\frac{1}{4\kappa_c} L[C]} \Bigl(
        2 + e^{-{1}/{\epsilon} }
        \sum_{s,s'\in C, s\not=s'} \>
	\langle e^{i(\varphi_0)_{s}} e^{-i(\varphi_0)_{s'}} \rangle
	+ \cdots \Bigr) \,.
    \label{eq:W_leading_eps}
\end{align}
The double sum in \eqref{eq:W_leading_eps}
has short-distance contributions, when $s$ and $s'$ are
close together relative to the loop perimeter $L[C]$, and long-distance
contributions when the separation of $s$ and $s'$ are of order
of the loop size.
The short distance contributions will be proportional to
the perimeter $L[C]$ and will exponentiate (when worldlines $\Xi$
comprising multiple line segments along $C$ are included),
leading to small $O(e^{-1/\epsilon})$ corrections to the screening mass,
\begin{align}
    m = \frac 1{4\kappa_c} + O(e^{-1/\epsilon}) \,.
\end{align}
For large loops, the long-distance contribution of the
above double sum will generate corrections proportional
to $L[C]^2$,
since the correlator
$
    \langle e^{i(\varphi_0)_{s}} e^{-i(\varphi_0)_{s'}} \rangle
$
approaches a constant, equal to the absolute
square of the (spontaneously broken)
vacuum expectation value
$
    v \equiv \langle e^{i \varphi_0} \rangle
$.
Consequently,
\begin{align}
    \langle W(C) \rangle = e^{-m L[C]} \,
    \bigl( \, 2+ e^{-{1}/{\epsilon}} L[C]^2 \, |v|^2 + \cdots \bigr) \,.
\end{align}
This correction cannot naively exponentiate (as doing so
would change perimeter law behavior into unphysical behavior
growing exponentially with $L[C]^2$).
Rather, it reflects the presence of two differing states
with slightly different values of the screening mass,
\begin{align}
    \langle W(C) \rangle
    =
    e^{-m_+ L[C] }
    +
    e^{-m_- L[C] } \,,
\end{align}
with
\begin{align}
    m_\pm \equiv
    \frac 1{4\kappa_c} \pm e^{-1/2\epsilon} \, |v|
    + O(e^{-1/\epsilon}) \,.
\end{align}
This shows that the long-distance corrections also
simply shift the screening mass
by a small $O(e^{-1/2\epsilon})$ amount,
with one of the two terms dominating as $L[C] \to \infty$.  

We now consider the vortex correlator
$\langle W(C) V(\tilde{\Gamma}) \rangle$.
The evaluation of this correlator
differs from the above treatment of
$\langle W(C) \rangle$ based on
\eqref{eq:W_leading_eps}
only through the change that the expectation value of
$e^{i \varphi_0}$ will now have a spatially
varying phase.
This will merely decrease the size of the (already small)
long-distance contribution to the double sum in
\eqref{eq:W_leading_eps},
leading to a completely analogous
result for the vortex-Wilson loop correlator,
\begin{align}
    \langle W(C)V(\tilde{\Gamma}) \rangle
    =
    \Bigl( e^{-m_+' L[C]} + e^{-m_-' L[C]} \Bigr) \,
    \langle V(\tilde\Gamma) \rangle \,,
\end{align}
with
$
    m_\pm' = (4\kappa_c)^{-1} \pm O(e^{-1/2\epsilon})
$.
Therefore in the confining regime under consideration,
\begin{align}
    \langle W(C) V(\tilde{\Gamma}) \rangle > 0 \,, 
\end{align}
showing that the magnetic flux carried by vortices is
$0 \textrm{ mod } 2\pi$.

\subsection{$\UB$-breaking Higgs regime}
\label{sec:U1_breaking_Higgs_regime}

Now let us suppose that $\kappa_{\pm} \gg 1$ as well as $\kappa_0 \gg 1$,
putting the theory into a regime where both $\varphi_0$ and $\varphi_\pm$
are all condensed.
But since $\varphi_\pm$ carries gauge charge, this more properly means
a Higgs regime.%
\footnote
    {%
    Gauge-charged operators like $\varphi_{\pm}$ cannot truly
    `condense' as a consequence of Elitzur's theorem~\cite{Elitzur:1975im}.
    }
We hasten to add that at this stage this is only a heuristic statement
as, a-priori, it is not clear how to sharply define a Higgs phase.

\subsubsection*{Strong gauge coupling: $\beta \ll 1$}

We will analyze the corners $\beta \ll 1$ and $\beta \gg 1$ separately,
and first consider the strong-coupling regime $\beta \ll 1$.
We will see that this domain is
smoothly connected to the continuum ``deep Higgs regime'' where
$\kappa_0$, $\kappa_\pm$, and $\beta$ are all large and,
e.g., the gauge field develops a parametrically large mass.  

In the strong-coupling regime, $\beta \ll 1$,
working with the worldvolume action \eqref{eq:gauge_worldsheet}
for the gauge field but leaving the matter fields in their
original form is most convenient.
The complete effective lattice action including insertions of
$W(C)$ and $V(\tilde{\Gamma})$ takes the form
\begin{align} \label{eq:Higgs_S_eff}
    S_{\textrm{eff}}(\Sigma,\varphi_0,\varphi_c,n_0,n_c,t;C, \tilde \Gamma)
    &=
   \frac{1}{2\beta} \sum_p \>
       ([\Sigma]_p-[D]_p)^2
   + \frac{\kappa_c}{2\kappa_+\kappa_-} \sum_\ell \>
       [\partial\Sigma]_\ell^2     
\nonumber \\ &
  + \frac{\kappa_0}{2} \sum_{\ell} \>
      [ (d\varphi_0)_\ell - 2\pi (n_0)_\ell ]^2
   + i \sum_{p}  \>
       \theta_{\star p} \, [(dn_0)_{p} -  (\star[\tilde{\Gamma}])_p ] 
\nonumber \\ &
    + \frac{\kappa_c}{2} \sum_\ell \> [(d\varphi_{c})_\ell-2\pi 
(n_c)_\ell ]^2
    + \frac{\epsilon}{2} \sum_s
	\left[(\varphi_0)_s - (\varphi_c)_s - 2\pi t_s \right]^2
\nonumber \\ &
    + 2\pi i \, \frac{ \kappa_c}{\kappa_-}\sum_{\ell} \>
	(n_c)_{\ell} \, [\partial\Sigma]_\ell 
      \, . 
\end{align}
The first two terms in this action are minimized when
$\Sigma = D$, so that $\partial\Sigma = C$,
with resulting perimeter-law behavior for the
Wilson loop,
$\langle W(C) \rangle \sim e^{-(\kappa_c/2\kappa_+\kappa_-) L(C)}$.
Large $\kappa_\pm$ combined with strong coupling, $\beta \ll 1$,
implies that screening by the charged fields is highly efficient.

The $\varphi_0,n_0$ terms on the second line of
\eqref{eq:Higgs_S_eff}, at large $\kappa_0$,
ensure (after integration over $\theta_{\star p}$) that
$n_0$ is non-zero only on a dual sheet ending on the vortex worldline $\tilde\Gamma$,
and that $\varphi_0$ jumps by $2\pi$ upon crossing this sheet.
Minimizing these terms leads to $\varphi_0$ having the expected
vortex profile with $|d\varphi_0-2\pi n_0|$ falling with
inverse distance away from the vortex worldline.
For a large but finite contractible vortex loop with perimeter
$L(\tilde\Gamma)$,
the resulting cost in action scales as
$L(\tilde\Gamma) \ln L(\tilde\Gamma)$, so that
$
    \langle V(\tilde\Gamma) \rangle
    \sim
    e^{-\kappa_0 \, C \, L(\tilde\Gamma) \ln L(\tilde\Gamma)}
$,
with $C$ an $O(1)$ constant dependent on the shape of
the vortex loop $\tilde\Gamma$.%
\footnote
    {%
    This estimate, and the following discussion, assumes
    that the vortex loop $\tilde{\Gamma}$
    has a single characteristic dimension
    (like a circle or square)
    so that the area of a minimal spanning surface scales
    as the square of the loop perimeter.
    }

Next, 
the $\varphi_c$-dependent terms in the third line of above action
will be minimized when $d\varphi_c \approx 2\pi n_c$
and simultaneously $\varphi_c \approx \varphi_0 \bmod 2\pi$.
For a large vortex loop $\tilde\Gamma$,
any configuration in which the winding of $\varphi_0$ 
on paths linking $\tilde\Gamma$
is not matched by corresponding winding in $\varphi_c$
(mod $2\pi$) will be highly suppressed,
with a cost in action growing at least as fast as the
area of a surface spanning $\tilde\Gamma$.
Consequently, the action is minimized when
$n_c = n_0$
and $\varphi_c$ and $\varphi_0$ have identical profiles
around the vortex,
so that these terms in the action are also of order
$L(\tilde\Gamma) \ln L(\tilde\Gamma)$.

The last term of the action \eqref{eq:Higgs_S_eff}
is imaginary and determines the phase of the
vortex-Wilson loop correlator.
The sum in this term,
for the minimal action configuration,
is just the linking number of the vortex and Wilson loops,
\begin{align}
    \sum_\ell \> (n_c)_\ell\, [\partial\Sigma]_\ell
    &= \sum_\ell (n_0)_\ell\, [C]_\ell 
   = \textrm{Link}(C,\tilde{\Gamma}) \,.
\end{align}
Therefore, in this strong coupling Higgs regime we find that
\begin{align}
    \langle W(C) V(\tilde\Gamma) \rangle
    &=
    e^{ 2\pi i \, (\kappa_c/\kappa_-)\, \textrm{Link}(C,\tilde{\Gamma}) }
    \, \langle V(\tilde\Gamma) \rangle \,,
\end{align}
showing that vortices and Wilson loops have non-trivial braiding
phases with each other.
In this gapless Higgs regime,
in the absence of $\ZF$ symmetry (i.e., when $\kappa_+ \ne \kappa_-$)
the Aharonov-Bohm phase of a vortex is not quantized
and depends on coupling constants of the theory,
in contrast to the more familiar examples of analogous braiding phases
in conventional topologically-ordered systems with a mass gap.
But in the $\ZF$ symmetric case where $\kappa_+ = \kappa_-$ and
$\kappa_c/\kappa_- = 1/2$,
the phase of $\langle W(C) V(\tilde\Gamma) \rangle$
does become quantized and equal to $\pm 1$,
as required by the earlier symmetry analysis, so that
\begin{align}
    \Omega &\equiv
    \frac{\langle W(C) V(\tilde\Gamma) \rangle}
    {|\langle W(C) V(\tilde\Gamma) \rangle|}
    = \exp \bigl[ i \pi \, \textrm{Link}(C,\tilde{\Gamma}) \bigr] .
\end{align}

\subsubsection*{Weak gauge coupling: $\beta \gg 1$}
\label{sec:continuum_limit_analysis}

We now consider the regime where $\kappa_0$, $\kappa_{\pm}$, and $\beta$
are all large compared to 1.
This is a semiclassical regime and is connected to the continuum
limit of our lattice theory.
One could analyze the physics in this regime by directly
minimizing the lattice action \eqref{eq:full_action},
with or without a vortex operator insertion.
But since we are only interested in long-distance properties 
and not details of, e.g., the vortex core structure,
it is far more convenient to use a continuum description.
To this end, let us introduce ``physical'' parameters by
defining $\kappa_0 = v_0^2 \, a_{\rm lat}$,
$\kappa_{\pm} = v_{\pm}^2 \, a_{\rm lat}$,
$\beta = 1/(e^2 \, a_{\rm lat})$, and
$\epsilon = \hat{\epsilon} \, a_{\rm lat}^3$.
The continuum description involves a $\U(1)$ gauge field $a$, charge $\pm 1$
$2\pi$-periodic Stueckelberg scalars $\varphi_{\pm}$, and a
$2\pi$-periodic Nambu-Goldstone boson field $\varphi_0$ with the
action
\begin{align}
    S_{\rm continuum} =
    \int d^3{x} \Biggl[
   \frac{1}{2e^2} (da)^2
   &+\sum_{\pm}  \frac{v_{\pm}^2}{2} \, (d\varphi_{\pm} \mp a)^2
   + \frac{v_0^2}{2} \, (d\varphi_0)^2
   - {\hat\epsilon} \, \cos(\varphi_0 {-} \varphi_{+} {-} \varphi_{-})
   + \cdots
    \Biggr] ,
\label{eq:Stueckelberg_continuum}
\end{align}
where $\cdots$ denotes higher harmonics in the Fourier
expansion of the original $\epsilon$-interaction.
In this description, a
unit-winding vortex operator $V(\tilde{\Gamma})$ can be defined as a defect
living on a closed curve $\tilde{\Gamma}$ in spacetime around which
$\varphi$ has monodromy $2\pi$.
Such a defect has a divergent (i.e., UV-cutoff sensitive) core action,
so the vortex operator may be regarded as a non-dynamical probe.

Consider a vortex loop for a curve $\tilde\Gamma$ whose radius of
curvature is larger than any microscopic scale of the theory,
so that in a large neighborhood around some point on $\tilde\Gamma$
(large compared to microscopic scales but small compared to the overall
size of $\tilde\Gamma$) the segment of $\tilde\Gamma$ passing through
this neighborhood is essentially straight.
Analyzing the behavior of the fields in \eqref{eq:Stueckelberg_continuum}
in this neighborhood is then a simple exercise, 
identical to the treatment in Ref.~\cite{Cherman:2020hbe}.
Taking cylindrical coordinates $(r,\theta, z)$ centered on
$\tilde{\Gamma}$, we can set $\varphi_0 = \theta$ to focus on the
case of a minimal vortex of winding number 1.
Minimizing the $\epsilon$ term contribution to the action then
requires that $\varphi_+$ have identical angular dependence
while $\varphi_-$ is constant, or vice-versa.
Outside the vortex core, the gauge field $a$ must rapidly
approach a flat connection, so by cylindrical symmetry
$a_\theta \sim \Phi/(2\pi r)$ for some total magnetic flux $\Phi$.
Minimizing the charged scalar kinetic terms then
determines the value of this flux.
One finds
\begin{align}
    \Phi = \frac{2 \pi \, v_+^2}{v_+^2 + v_-^2}
    = \frac{2\pi \kappa_c}{\kappa_-}
    \,,
\end{align}
if $v_+^2 \le v_-^2$ (with $v_+ \leftrightarrow v_-$ otherwise).

A\ Wilson loop $W(C)$ encircling the vortex acquires
an Aharonov-Bohm phase which simply measures this
magnetic flux, so to leading order in this weak-coupling regime,
\begin{align}
    \langle W(C) V(\tilde\Gamma) \rangle
    =
    \Omega \, \langle V(\tilde\Gamma) \rangle \,,
\end{align}
with
\begin{align}
    \Omega = e^{i\Phi} = e^{2\pi i \kappa_c / \kappa_-} \,.
\end{align}
A\ Wilson loop which does not link the vortex,
and is always well outside the vortex core,
acquires no Aharonov-Bohm phase.

This demonstrates that deep in the 
$\kappa_0, \kappa_{\pm} \gg 1$ Higgs regime,
Wilson loops and vortex loops have the same linking-dependent
braiding phases for weak gauge coupling, $\beta \gg 1$, as they do
at strong coupling, $\beta \ll 1$.
In fact, there is a simple argument that implies that these two regimes
should be smoothly connected.
Following Ref.~\cite{Banks:1979fi}, consider taking $\kappa_\pm \to
\infty$.
Then the charged scalar kinetic terms
in the lattice action \eqref{eq:full_action}
become constraints requiring that
\begin{equation}
    a_\ell = (d\varphi_+)_\ell - 2\pi (n_+)_\ell
    =  - (d\varphi_-)_\ell + 2\pi (n_-)_\ell\,. 
\end{equation}
This implies that $(da)_p = -2\pi (dn_+)_p = 2\pi (dn_-)_p$
so that the part of the lattice action involving the gauge field
reduces to
\begin{equation}
    2 \pi^2 \beta \sum_p \left[ (dn_-)_p - m_p \right]^2 \,.
\end{equation}
Now one can perform the field redefinition $m_p \to m_p + (dn_-)_p$, 
integrate over $\varphi_\pm$
(which contributes a $\beta$-independent overall
factor in the path integral) and arrive at
\begin{equation}
    S_{\kappa_\pm \to \infty}
    = 2\pi^2 \beta \sum_p m_p^2 + S_{\text{neutral}}\,. 
\end{equation}
The trivial sum over $m_p$ configurations can then be performed, yielding
an overall factor in the path integral which is a smooth analytic
function of $\beta$.
Therefore,
there is no phase transition as a function of $\beta$ in
the strict $\kappa_\pm \to \infty$ limit.
And consequently, any putative non-analyticity must lie in the interior
of the phase diagram at some finite values of $\kappa_\pm$,
with smooth continuity between the weak and strong coupling regimes
assured above these finite values.
It thus makes sense to regard the entire
$\kappa_0, \kappa_{\pm}\gg 1$ regime, for any value
of $\beta$, as a single $\UB$-broken Higgs phase.

\section{Phase diagram}
\label{sec:phase_diagram}

We now discuss the implications of our results for the
phase diagram of our model.
As noted earlier, this theory
has a spontaneously broken $\UB$
symmetry throughout our parameter region of interest where 
$\kappa_0$, $\kappa_{\pm}$, $\beta$, and $\epsilon$
are all positive. To simplify
the discussion we will focus on the parameter slice $\kappa_{+} =
\kappa_{-}$, where the theory also enjoys a $\ZF$ symmetry.
Then, as long as the $\ZF$ symmetry is not spontaneously broken,
our Aharonov-Bohm phase $\Omega$
defined via a correlation function of line operators
is quantized and must equal either $+1$ or $-1$.
This was verified explicitly in the tractable periphery
of the phase diagram,
including both confining and Higgs regimes.
We found that $\Omega = +1$ deep in a
calculable corner of the confining regime, but $\Omega = -1$ deep in the
calculable corners of the Higgs regime.  Since the phase of $\Omega$ is quantized,
this means that it must change abruptly (and non-analytically) along
some curve in the $\beta-\kappa_c$ plane.
This, of course, suggests
that $\Omega$ is a natural gauge invariant
observable that can provide a sharp distinction between
Higgs and confining regimes, as
illustrated in Fig.~\ref{fig:phase_diag}.

The fundamental question is whether the red ``transition'' line in
Fig.~\ref{fig:phase_diag} necessarily
denotes a genuine bulk phase transition.
If it does not, then the question is why not -- how might non-analyticity
in $\Omega$ be understood if this behavior is not associated with
non-analyticity in ground state properties, i.e., a thermodynamic phase transition?

One might wonder if 
$\Omega$ could cease to be well-defined everywhere along the red curve  in
Fig.~\ref{fig:phase_diag} due to a divergence in the vortex core size.%
\footnote
    {%
    Our $\ZF$ symmetry
    forbids mixing between the photon and the Nambu-Goldstone boson, so the magnetic
    flux of $V(\tilde{\Gamma})$ is exponentially well-localized (with a scale set by
    the Meissner mass) near $\tilde{\Gamma}$; see the appendix of
    Ref.~\cite{Cherman:2020hbe} for an explicit analysis. \label{fn:core_size}
    }
However, a divergent vortex core size (or vanishing Meissner mass)
would itself signal a bulk phase transition due to the appearance of
an additional gapless mode \emph{in the bulk}.
It is important to distinguish this from the distinct phenomenon known to
occur on defects with $d>1$ worldvolumes where there can be a second-order phase
transition on the defect, so that new gapless excitations appear on the defect
worldvolume while the bulk remains gapped, see e.g.,
Refs.~\cite{Gaiotto:2017tne,Komargodski:2017smk,Anber:2023urd}.  A diverging
vortex core size would reflect a divergence in the correlation length of some
mode in a spatial volume whose transverse size from the defect is itself diverging.
That is fundamentally different and signals the appearance of a new bulk gapless
mode, which would in turn imply a bulk phase transition due to the usual
non-analyticities associated with new modes going gapless on subdomains of the
phase diagram.

So putting aside this hypothetical possibility of a diverging vortex core size,
another possibility is that the vortex phase
\begin{equation}
    \Omega \equiv
     \frac{\langle W(C) V(\tilde\Gamma) \rangle}
     {|\langle W(C) V(\tilde\Gamma) \rangle|}
\end{equation} 
might be non-analytic, without a bulk phase transition, if at some point
it ceases to be well-defined due to line operator correlator
$\langle W(C) V(\tilde\Gamma) \rangle$ passing through zero.
This could happen if there are contributions to the
correlator from both vortices with flux $\pi$ (mod $2\pi$)
and vortices with flux zero (mod $2\pi$), 
with the magnitudes of these two contributions crossing 
on some curve in parameter space, thus producing an
exact cancellation in the 
$\langle W(C) V(\tilde\Gamma) \rangle$ correlator.%
\footnote
    {%
    The precise position in the phase diagram of such a hypothetical cancellation
    would also depend on the specific shape of the vortex worldline $\tilde\Gamma$,
    since changes in this curve would differentially perturb the magnitude of the
    two vortices' contributions to $\langle V(\tilde\Gamma) \rangle$
    due to the (generically) differing core energies of the two types of vortices.
    } 
In other words, a level-crossing phenomena between flux-0 and flux-$\pi$
vortices could cause non-analyticity in $\Omega$.
    
Is this plausible?
We cannot rule this scenario out using the tools at our disposal.
Our calculations for the vortex states are reliable deep in the
confining regime and the Higgs regime.
Within these calculable regimes, a single type of vortex accounts for the
behavior of the vortex expectation value $\langle V(\tilde\Gamma)\rangle$
and associated $\langle W(C) V(\tilde\Gamma) \rangle$ correlator.
We see no evidence of any locally metastable vortex states.
However, one can imagine that such metastable states appear
only in intermediate regions of parameter space where none of our
semiclassical calculations are reliable.
Then these putative coexisting vortex states may level-cross,
causing the $\langle V(\tilde\Gamma) W(C) \rangle$ correlator to pass through zero.
If this is the case, then in the parameter space
around the level-crossing line the higher energy vortex
state must be unstable to decay to the minimum energy vortex.
In our previous work~\cite{Cherman:2020hbe},
we noted that if the higher-energy ``wrong flux''
vortex were to decay to the lower-energy vortex via some sort
of instanton process on the vortex world sheet --- i.e., via a process which is
well localized on the vortex worldsheet --- this would imply the existence of 
a vortex junction carrying fractional magnetic flux in violation of
Dirac flux quantization.
In other words, such a world-sheet localized decay process is impossible.
However, just as a metal transitioning into a superconducting state can
simply expels a portion of an applied magnetic field whose flux is a
non-integer multiple of the magnetic flux quantum,
the putative higher-energy vortex in the above
level-crossing scenario can shed (or absorb) flux and decay into the lower
energy vortex state by producing an outward moving shell of radiation which
carries the ``unwanted'' flux.
Within the semiclassically tractable Higgs regime, one may explicitly
verify that this decay process of is energetically feasible.

Where does this leave us?
On the upper edge of the
schematic phase diagram of Fig.~\ref{fig:phase_diag},
the limit $\beta\to\infty$ effectively turns off the gauge field and theory
acquires an extra global $U(1)$ 0-form symmetry
and should lie in the same universality class as the 3D XY-model,
with a continuous phase as a function of $\kappa_\pm$ as in the
classic analysis of Ref.~\cite{Fradkin:1978dv}.
Moving inward from this XY transition should be a line of first
order transitions,
as may be seen in a reliable weak-coupling perturbative analysis
of 3D gauge-Higgs systems, see e.g. Ref.~\cite{Cherman:2020hbe}.
In the asymptotically large-$\beta$, large-$\kappa_{\pm}$ regime, it is clear that this
bulk transition line coincides with the line at which $\Omega$
changes sign. 

One viable possibility is that this transition line
extends all the way to $\beta = 0$
(perhaps encountering some intermediate spontaneously broken
$\ZF$ symmetry zone along the way, as indicated in Fig.~\ref{fig:phase_diag}).
Alternatively, this transition line might terminate
in a critical point beyond which a level-crossing line emerges across which
$\Omega$ changes sign.
Should this scenario take place, it would be
quite interesting in its own right since it would involve the simultaneous
existence of two types of vortices but only in a deeply quantum regime.
Numerical simulations should be able to reveal whether or not 
vortices carrying different fluxes can meaningfully coexist in our model,
and should be able to map out the phase diagram and the extent to which
thermodynamic phase transitions coincide with changes in $\Omega$.

To prove the existence of a phase boundary, and rule out the level-crossing scenario,
perhaps one could show that the
Higgs regime is some sort of gapless symmetry protected topological (SPT) or symmetry
enriched topological (SET) phase, characterized by some non-trivial
quantized topological term (involving either background or dynamical fields)
in the long-distance effective action.%
\footnote
    {%
    For instance, to look for possible SPT phases one could imagine
    turning on background fields for the $(\ZZ_2)_F$ symmetry, which
    is not involved in any 't Hooft anomalies (thereby ensuring that
    possible SPT terms could not be absorbed by field redefinitions,
    see e.g. Sec. 2.5 in Ref.~\cite{Dumitrescu:2023hbe} for a summary
    of these issues). However, the partition function in the presence
    of $(\ZZ_2)_F$ background fields is positive for all values of
    the couplings, ruling out a non-trivial SPT response. One could
    also turn on backgrounds for the unbroken $U(1)^{(1)}$ and
    charge conjugation symmetries, but anomaly considerations make
    this analysis more subtle.
    } 
Unfortunately gapless
topological phases are far less well understood than gapped topological phases
of matter, and we have not succeeded in finding such an explanation.%
\footnote
    {%
    For some related attempts  in this direction, see
    Refs.~\cite{Hirono:2018fjr,Hidaka:2019jtv,Hidaka:2022blq}, which discuss
    whether topological order can appear in Higgs phases with fundamental
    representation matter fields. Conventional topological order can be
    interpreted in terms of spontaneous breaking of a higher-form symmetry.
    Hirono and Tanizaki identified a putative emergent
    higher-form symmetry which is not
    spontaneously broken, and interpreted this
    result as an argument in favor of quark-hadron continuity
    \cite{Hirono:2018fjr}.
    However,
    as discussed in Ref.~\cite{Cherman:2023xok}
    there are reasons to be skeptical that the
    ``emergent symmetry'' contemplated in
    Ref.~\cite{Hirono:2018fjr} is well-defined.
    Moreover, we do not agree with the starting
    premise of Ref.~\cite{Hirono:2018fjr} that the only way to argue for a
    phase transition in the context of our results here and in
    Refs.~\cite{Cherman:2018jir,Cherman:2020hbe} is to relate
    the change in vortex phase to the
    spontaneous breaking of a higher form symmetry,
    which would gives rise to an essentially standard
    sort of topological order.  }

Finally, as noted earlier, one could simply study our lattice model
\eqref{eq:full_action} numerically.
This model is well-suited for exploration using Monte Carlo simulations.
It should be quite feasible to perform simulations 
scanning the physics in the $\beta$-$\kappa_c$ plane and
testing whether changes in $\Omega$ are associated with
bulk phase transitions. If a phase boundary exists, it would persist all the way down to $\beta =0$, which is a natural starting point for numerical studies. In this infinite gauge coupling limit, integrating out the gauge field binds the charged fields $\varphi_\pm$ into a single composite $\varphi_c = \varphi_+ + \varphi_-$ described by a simple effective action,
\begin{align}
S_{\beta = 0} &= \frac{\kappa_0}{2}\sum_\ell \left[ (d\varphi_0)_\ell - 2\pi (n_0)_\ell \right]^2  + i \sum_p \theta_{\star p}(dn_0)_p
\nonumber\\
&+ \frac{\kappa_c}{2}\sum_\ell \left[ (d\varphi_c)_\ell - 2\pi (n_c)_\ell \right]^2 + \frac{\epsilon}{2}\sum_s \left[ (\varphi_0)_s - (\varphi_c)_s - 2\pi t_s \right]^2\,.
\end{align}  
In this limit, the Wilson loop reduces to 
\begin{subequations}
\begin{align}
    W(C)\Big|_{\beta = 0} &=
    e^{-\frac{\kappa_c}{2\kappa_+\kappa_-}L(C)} \, \exp\bigg(\!-2\pi i \frac{\kappa_c}{\kappa_-} \sum_{\ell\in C} (n_c)_\ell \bigg)
\\ &
    \stackrel{\kappa_+ = \kappa_-}{\longrightarrow}
    e^{-\frac{1}{4\kappa_\pm} \, {L(C)}} \, \exp\bigg(\!-\pi i  \sum_{\ell\in C} (n_c)_\ell \bigg) \,.
\end{align}
\end{subequations}
and the order parameter $\Omega$ probes whether inserting a vortex for $\varphi_0$ induces a vortex for $\varphi_c$. We hope to perform such numerical studies in the future.

\section{Conclusions}
\label{sec:conclusions}

In this paper we have sharpened the proposal in our earlier papers
Refs.~\cite{Cherman:2020hbe,Cherman:2018jir} that,
within ``superfluid'' (or spontaneously broken $U(1)$) phases),
the Aharonov-Bohm phase $\Omega$ of
charged particles encircling a minimal vortex can be used to
distinguish the Higgs and confinement regimes of gauge theories,
despite the absence of any distinguishing local order parameters.
We expressed $\Omega$ in terms of a standard correlation function of two
line operators,
\begin{equation}
    \Omega = \exp\left[ i \arg\, \langle W(C) V(\tilde\Gamma) \rangle \right] =
    \frac{\langle W(C) V(\tilde\Gamma) \rangle}
    {|\langle W(C) V(\tilde\Gamma) \rangle|} \,,
\label{eq:Omegafinaldef}
\end{equation} 
where the minimal distance between contours $C$ and $\Gamma$ is large compared to the
vortex core size.  A key virtue of our lattice construction is that it allowed
us to compute $\Omega$ in the confining regime with string-breaking effects
fully taken into account. This is something which could not be done
quantitatively in previous continuum approaches
\cite{Cherman:2020hbe,Cherman:2018jir}. We took advantage of lattice dualities
and strong coupling expansions to compute $\Omega$ explicitly in Higgs and
confining regimes and found that
\begin{align}
    \Omega = \begin{cases}
        +1\,, & \textrm{confining regime}; \\
        -1\,, & \textrm{Higgs regime} \,.
    \end{cases}
\end{align}
We discussed whether this non-analytic change in $\Omega$ might
result from a level-crossing phenomena without signaling a genuine thermodynamic
phase transition separating Higgs and confining regimes.
However, at present we can neither demonstrate nor exclude such a
level-crossing scenario as the origin of non-analyticity in $\Omega$.%
\footnote
    {%
    We have attempted, unsuccessfully, to engineer symmetry-preserving deformations
    of our model which could possess multiple locally stable vortices with differing
    magnetic flux within a semiclassically tractable regime.
    }
A numerical or analytic demonstration of the coexistence of vortices with
distinct fluxes would thus be of independent interest.  In any case, the natural
next step from the analysis in this paper is perform numerical simulations
of the model we have defined, studying its phase diagram and vortex correlation
functions.

A major caveat for phenomenological applications is that our definition 
of $\Omega$ \eqref{eq:Omegafinaldef}
is only well-defined in theories, such as the model examined in this paper,
possessing a $1$-form symmetry that acts on vortex operators
$V(\tilde\Gamma)$.
The benefit of using expression \eqref{eq:Omegafinaldef}
to define $\Omega$ is that it is
completely concrete, is not tied to any semiclassical approximations,
and should be straightforward to measure in lattice simulations.
We believe that doing so would be very worthwhile.
It would, of course, be desirable to understand to what extent our conclusions
can also apply to superfluid phases in $d$-dimensional theories which do
not have exact $(d-2)$-form symmetries acting on vortices.
In particular, high density QCD lacks such a symmetry.
There is a sense in which a spontaneously-broken
$U(1)^{(0)}$ symmetry leads to an emergent  $U(1)^{(d-2)}$ symmetry
in the long distance limit
\cite{Delacretaz:2019brr,Cherman:2023xok},
but the precise details are subtle.
In such theories, 
the presence of dynamical vortices can make it difficult
to isolate properties of a single vortex, and one cannot define an
order parameter --- a vacuum correlation function --- in the same
manner as done in this paper.  One way to avoid this issue would
be to study finite rotating ``superfluid'' systems
which will have stable unscreened vortices.
This, of course, is of direct relevance
to neutron star physics.
More generally, it seems likely that sufficiently dilute dynamical vortex
excitations will be unable to turn a bulk phase transition into a smooth crossover.
In particular, in three spacetime dimensions
adding dynamical vortices is equivalent to adding massive particles
with unit charge under the emergent gauge field on the gauge theory
side of particle-vortex duality
(as discussed in Appendix~\ref{sec:AH_appendix}).
If there is a phase boundary when these particles are infinitely
massive, then it should persist when they are very heavy compared
to the energy scales associated with
the phase transition, which should be the case in many applications.

\section* {Acknowledgments}

We are grateful to M. Alford, T. Dumitrescu, J. Greensite, S. Moroz, N. Seiberg,
and T. Sulejmanpasic for useful discussions over the last several years. We acknowledge support from the
Simons Foundation through the Collaboration on Confinement and QCD Strings under
award number 994302 (AC), 
as well as from the U.S.~Department of Energy through the grant
DE-SC\-0011637 (LY) and its Nuclear Physics Quantum Horizons program through the
Early Career Award DE-SC0021892  (SS). TJ acknowledges support from a Doctoral Dissertation Fellowship from the University of Minnesota, as well as the Mani L.~Bhaumik Institute for Theoretical Physics at UCLA. 

\appendix

\section{Differential forms on the lattice}
\label{sec:lattice_conventions}

We work on a cubic lattice with periodic boundary conditions. We have sites $s$
(or 0-cells $c^{(0)}$), links $\ell$ (or 1-cells $c^{(1)}$), plaquettes $p$ (or
2-cells $c^{(2)}$), cubes $c$ (or 3-cells $c^{(3)}$), etc. Each of these
$r$-cells carries an orientation, which when reversed takes $c^{(r)} \to
-c^{(r)}$. We also have the dual lattice, with the dual cells indicated by
tildes: the dual of an $r$-cell $c^{(r)}$ is denoted by $(\star c)^{(d-r)} =
\tilde c^{(d-r)}$. For example, in three dimensions $\star x = \tilde c$, $\star
\ell = \tilde p$, etc. The square of the duality operation on $r$-cells
satisfies $\star^2 = (-1)^{r(d-r)}$. We will mostly work in three dimensions
where this sign is always 1. Curves, surfaces, etc. are represented by linear
combinations of $r$-cells. 

It is helpful to pick a convention for the `positive' orientation when referring
to $r$-cells. Then in formulas we just sum over the positive orientation for
each $r$-cell, accounting for the negative orientation with minus signs.

Differential $r$-forms are linear functions on $r$-cells which satisfy
$\omega_{-c^{(r)}} = -\omega_{c^{(r)}}$. We can `integrate' an $r$-form on a
collection of $r$-cells $\Sigma^{(r)}$ by extending $\omega_{c^{(r)}}$ linearly
to $\Sigma^{(r)}$. A simple example of an $r$-form is the characteristic
function $[\Sigma]_{c^{(r)}}$ of a set $\Sigma$ of $r$-cells, which is equal to
the number of oriented times $c^{(r)}$ is traversed by $\Sigma$. 

Dual forms are defined very simply via 
\begin{equation}
(\star\omega)_{\tilde c^{(d-r)}} = \omega_{c^{r}}\,,
\end{equation}
so for example $(\star [C])_{\tilde p}$ is the oriented number of times the curve $C$ pierces the plaquette $\tilde p$ on the dual lattice. The exterior derivative is defined by
\begin{equation}
(d\omega)_{c^{(r+1)}} = \sum_{c^{(r)} \in \partial c^{(r+1)}} \omega_{c^{(r)}}\,.
\end{equation}
We also define the divergence operator $\delta = \star d \star$ which takes an $r$-form to an $(r-1)$-form. We can integrate $r$-forms on various submanifolds $\Sigma^{(r)}$ by using the characteristic function, 
\begin{equation}
\sum_{c^{(r)} \in \Sigma} \omega_{c^{(r)}} = \sum_{c^{(r)}} [\Sigma]_{c^{(r)}} \omega_{c^{(r)}} = \sum_{c^{(r)}}  \omega_{c^{(r)}}\,(\star [\Sigma])_{\tilde c^{(d-r)}}
\end{equation}
where the latter sum is over \emph{all} $r$-cells on the lattice. This is equivalent to the continuum formula
\begin{equation}
\int_{\Sigma^{(r)}} \omega^{(r)} = \int \omega^{(r)} \wedge \delta^{(d-r)}(\Sigma^{(r)})
\end{equation}
More generally given two $r$-forms $a_{c^{(r)}}$ and $b_{c^{(r)}}$, 
\begin{equation}
\sum_{c^{(r)}} a_{c^{(r)}} b_{c^{(r)}}  \ \sim \ \int a \wedge \star b = \int b \wedge \star a. 
\end{equation}
Similarly, we can `integrate' by parts: 
\begin{subequations}
\begin{align}
\sum_{c^{(r)}} (d\omega)_{c^{(r)}} b_{c^{(r)}} =\sum_{c^{(r)}} (d\omega)_{c^{(r)}} (\star b)_{\tilde c^{(d-r)}} &= (-1)^r \sum_{c^{(r-1)}} \omega_{c^{(r-1)}} (d\star b)_{\tilde c^{(d-r+1)}} 
\label{eq:sumbypartsA}\\
&= (-1)^r \sum_{c^{(r-1)}} \omega_{c^{(r-1)}} (\delta b)_{c^{(r-1)}}
\end{align}
\end{subequations}

\section{Review of particle-vortex duality on the lattice}
\label{sec:AH_appendix}
To understand how one inserts a global vortex in, for example, the XY model, it is helpful to consider particle vortex duality. The XY model is 
dual to the Abelian-Higgs theory of charge $1$. The duality maps charge-$1$ Wilson lines in the Abelian-Higgs model to a global vortex on 
the XY side. Similarly, a particle excitation world-line in the XY model maps to a gauged vortex excitation on the Abelian-Higgs side.  Since, 
the insertion of a Wilson line on the Abelian-Higgs side is standard,
this will be our starting point. The application of duality will then reveal the appropriate definition of a global 
vortex insertion on the XY side.
We will also implement the reverse procedure for the gauged
vortex on the Abelian-Higgs side, i.e. beginning with 
a particle world line on the XY side and then follow the duality to reveal the insertion of a gauged Abelian-Higgs vortex. 

To keep our discussion somewhat general we begin with a $U(1)$ gauge theory coupled to a charge-$N$
scalar field. We will use a Villain formulation for the lattice
action. The action of this theory, formulated on the dual lattice,
is given by
\begin{align}
    \label{eq:Abelian_higgs_no_monopoles}
    S_{\textrm{AH}} =
    \sum_{\tilde p}
	\frac{1}{2(2\pi)^2\kappa} \,
	[(d b)_{\tilde p} - 2\pi s_{\tilde p}]^2
    - i\sum_{x} \varphi_x (ds)_{\tilde c} 
    + \sum_{\tilde\ell} \frac{\lambda}{2N^2} \,
	[ (d \chi)_{\tilde\ell} - N  b_{\tilde\ell} - 2\pi v_{\tilde\ell} ]^2\,.
\end{align}
Here $\kappa$ and $\lambda$ are positive
parameters, and $s, v \in \ZZ$, $\varphi, b, \chi\in \mathbb{R}$ are 
fields to be summed or integrated. Integration of the Lagrange
multiplier field $\varphi$ ensures that there are no dynamical magnetic
monopole-instantons.  
The above action has the gauge redundancy
\begin{align}
    b_{\tilde\ell} &\to b_{\tilde\ell} + (d\omega)_{\tilde\ell} + 2\pi m_{\tilde\ell}, \, s_{\tilde p} \to s_{\tilde p} + (dm)_{\tilde p}\,,
\nonumber \\
    \chi_{\tilde x} &\to \chi_{\tilde x} + N\omega_{\tilde x} + 2\pi h_{\tilde x}, \, v_{\tilde\ell} \to v_{\tilde\ell} + (dh)_{\tilde\ell} - N m_{\tilde\ell} \,,
\label{eq:gaugetransformations11} 
\\
    \varphi_x &\to \varphi_x + 2\pi k_x\,, 
\nonumber
\end{align}
where $\omega \in \RR$, $m,h,k \in \ZZ$.

Performing a Poisson resummation on $s$, followed by the Gaussian
integral over $b$, and dropping total derivatives, yields
another description of the same model involving the scalar field
$\varphi_x$:
\begin{align}
    S_{\varphi} = \sum_{\ell} \frac{\kappa}{2} \, [ (d\varphi)_\ell - 2\pi n_\ell ]^2 
     +\sum_{p} \frac{1}{2\lambda} \, (dn)_p^2 
     - i \sum_{p} \frac{2\pi}{N} \, v_{\star p} (dn)_{p}  \,,
    \label{eq:dual_vortex}
\end{align}
where $n_\ell \in \ZZ$ is an ``emergent'' integer field Poisson dual to
$s_{\tilde\ell}$.
This field
transforms as $n_\ell \to n_\ell + (dk)_\ell$ under the
gauge symmetry \eqref{eq:gaugetransformations11}. 
In addition to the gauge redundancy making the scalar field $\varphi$ compact,
the exponential of the action above is also invariant under the 
transformation
\begin{align}
v_{\tilde\ell} \to v_{\tilde\ell} + (dh)_{\tilde\ell} - N m_{\tilde\ell}, 
\end{align}
thanks to the summation-by-parts identity \eqref{eq:sumbypartsA}
together with the fact that $d^2 =0$.
Finally, $S_\varphi$ enjoys the global symmetry
\begin{align}
\UB: \; (\varphi_{0})_s \to (\varphi_{0})_s + \alpha, \quad \alpha \in [0,2\pi) \,,
\end{align}
and $e^{i\varphi_{0}}$ is a well-defined local operator that carries unit $U(1)$ charge.
The transformation from $S_{\textrm{AH}}$ to $S_{\varphi}$ is
an exact lattice realization of particle-vortex duality. 

We can now introduce Wilson loops in the Abelian-Higgs side $W_w(\tilde{\Gamma})$ in order to understand how they map to vortex insertions
on the dual theory. Tracking through the duality with an insertion of a
Wilson loop, we find that inserting $W_w(\tilde{\Gamma})$ into the path integral is
equivalent to replacing the right-most two terms in
\eqref{eq:dual_vortex} by
\begin{align} 
     \sum_{p} \frac{1}{2\lambda} \, \left[(dn)_p-w[\tilde \Gamma]_{\star p}\right]^2 
     - i \sum_{p} \frac{2\pi}{N} \, v_{\star p} \left[(dn)_p-w[\tilde \Gamma]_{\star p}\right]  \,.
\end{align}
The shift of the second term is equivalent to inserting the vortex operator \eqref{eq:vortex_operator} while the first term is an additional modification analogous to \eqref{eq:other_vortex_operator} which disappears in the limit $\lambda \to \infty$. 

\section{The $U(1)^{(0)}\times U(1)^{(1)}$ anomaly on the lattice}
\label{sec:anomaly_appendix}

Consider the modified Villain action \eqref{eq:villain_scalar_1form} for a compact scalar without dynamical vortices, 
\begin{align} 
    S = 
    \frac{\kappa}{2} \sum_{\ell} \> [ (d\varphi)_\ell - 2\pi n_\ell ]^2  
    - i\sum_{\tilde{\ell}} \theta_{\tilde{\ell}} \> (dn)_{\star \tilde{\ell}}  \,.
\end{align}

We use the Einstein summation convention for lattice indices to reduce clutter. This model has two interesting global symmetries, the 
$U(1)^{(0)}$ shift symmetry and the $U(1)^{(1)}$ vortex symmetry. They can be associated to the currents
\begin{subequations}
\begin{align}
J^{(1)}_\ell &= -i \kappa \left[ (d\varphi)_\ell - 2\pi n_\ell \right] ,\\ 
J^{(2)}_{\tilde p} &= \frac{1}{2\pi}\left[ (\star d\varphi)_{\tilde p} - 2\pi (\star n)_{\tilde p}\right]. 
\end{align}
\end{subequations}
The $U(1)^{(0)}$ symmetry of the model defined via
\eqref{eq:villain_scalar_1form} is
generated by%
\footnote
    {%
    The expression for $U_{\alpha}(\tilde\Sigma)$ in
    \eqref{eq:shift_sym_generator} is only valid up to contact terms
    in correlation functions where multiple $U_\alpha$ operators collide. To write an expression with an equal sign requires
    choosing some counter-terms.  Alternatively, one can avoid the need to specify
    counter-terms by using the Noether current following from a first-order form of the
    action instead of the one quoted in the main text.
    }
\begin{align}
    U_{\alpha}(\tilde\Sigma)
\sim \exp\bigg(i \alpha  \sum_{\tilde p\in \tilde\Sigma } (\star J^{(1)})_{\tilde{p}}  \bigg) ,
\label{eq:shift_sym_generator}
\end{align} 
while the $U(1)^{(1)}$ symmetry is generated by 
\begin{align}
    U_{\beta}(C)
    = \exp\bigg(i \beta  \sum_{\ell\in C}  (\star J^{(2)})_{\ell} \bigg),
\end{align}
where $\tilde \Sigma$ is a closed surface on the dual lattice
while $C$ is a
closed curve on the original lattice.

One may couple external gauge fields to these currents and look for anomalies. The
background gauge field coupling to the $U(1)^{(0)}$ current consists of
a pair $(A_\ell, X_p)$ where $A_\ell$ is $\RR$-valued and $X_p$ is
$\ZZ$-valued.
The background 1-form gauge redundancy acts as $A_\ell
\to A_\ell + 2\pi K_\ell$, $X_p \to X_p + (dK)_p$ and ensures that
this pair describes a compact $U(1)$ gauge field. Similarly, the 2-form background gauge field which couples to the $U(1)^{(1)}$ current consists of a pair
$(B_{\tilde p}, Y_{\tilde c})$ with $B_{\tilde p} \in \RR$, $Y_{\tilde c} \in \ZZ$,
and with the gauge redundancy
$B_{\tilde p} \to B_{\tilde p} + (d\beta)_{\tilde p} + 2\pi L_{\tilde p}$, $Y_{\tilde c} \to Y_{\tilde c} + (dL)_{\tilde c}$. 

The complete action of our scalar field $\varphi$ coupled to these gauge fields reads 
\begin{align} \label{eq:bgdaction1} 
    S &= \frac{\kappa}{2}\sum_\ell \left[ (d\varphi)_\ell -A_\ell - 2\pi n_\ell \right]^2
    -i\sum_p \theta_{\star p} \left[(dn)_p+X_p\right]
\nonumber\\ &\qquad{}
    - \frac{i}{2\pi }\sum_p B_{\star \ell} \,
	[(d\varphi)_{\ell} - A_{\ell} -2\pi  n_{\ell}]
    -i \sum_s  \varphi_s \, Y_{\star s} \,.
\end{align}
The full gauge invariance we would like to impose is
\begin{subequations}
\begin{align}
    \varphi_s &\to \varphi_s + \alpha_s + 2\pi k_s \,,
    &n_\ell &\to n_\ell + (dk)_\ell - K_\ell \,,
\\
    A_\ell &\to A_\ell + (d\alpha)_\ell + 2\pi K_\ell \,,
    &X_p &\to X_p + (dK)_p \,,
\\
    \theta_{\tilde\ell} &\to \theta_{\tilde\ell} + \beta_{\tilde\ell} \,,
    &B_{\tilde p} &\to B_{\tilde p} + (d\beta)_{\tilde p}  + 2\pi L_{\tilde p} \,,
\end{align}
\end{subequations}
with $K, L \in \ZZ$. While the last term in \eqref{eq:bgdaction1} looks somewhat unnatural, it is crucial to ensure that the background gauge variation of the action does not involve terms with \emph{dynamical} fields. The gauge variation of Eq.~\eqref{eq:bgdaction1} gives 
\begin{align}
    \Delta S
    &=
    -i \sum_p \beta_{\star p}[ (dn)_p + X_p ]
    - \frac{i}{2\pi}\sum_\ell  \, [(d\beta)_{\star \ell}+ 2\pi L_{\star\ell} ]
	[(d\varphi)_{\ell} -A_{\ell} - 2\pi  n_{\ell} ]
\nonumber\\ & \hspace*{1.6in}
    - i\sum_s \, \alpha_s[Y_{\star s}+ 2\pi (dL)_{\star s}]+ \varphi_s (dL)_{\star s}
\nonumber\\
    &= \frac{i}{2\pi}\sum_p  \beta_{\star p}[ (dA)_p - 2\pi X_p]
    + i \sum_\ell L_{\star\ell} \, A_{\ell}
    - i \sum_s \alpha_s[Y_{\star s}+ (dL)_{\star s}] \,,
\label{eq:anomaly1}
\end{align}
where we dropped total derivatives and integer multiples of $2\pi i$. This anomalous gauge variation consists of mixed bilinears of gauge transformations and background gauge fields, with at most 
one derivative. We can try to cancel the anomaly using local counterterms. The only relevant counterterm is $A_\ell B_{\star\ell}$; if we add it to the action with coefficient $-i/2\pi$ we can cancel some of the terms in \eqref{eq:anomaly1}, since
\begin{align}
\Delta (-\frac{i}{2\pi} \sum_\ell A_\ell B_{\star\ell}) &= -\frac{i}{2\pi}\sum_p \beta_{\star p} (dA)_{p}-i \sum_\ell L_{\star\ell }A_{\ell}
\nonumber\\
 & + \frac{i}{2\pi}\sum_s \alpha_s 
[(dB)_{\star s} + 2\pi (dL)_{\star s}] - i\sum_{\ell} K_{\ell} (B_{\star\ell} + (d\beta)_{\star\ell}) \,,
\end{align}
mod $2\pi i$. So we can cancel some of the terms in Eq.~\eqref{eq:anomaly1} at the cost of introducing other gauge-non-invariant terms, and the 
gauge variation becomes
\begin{equation}
\Delta S + \Delta S_{\text{c.t.}} = -i\sum_p \beta_{\star p}(X_p+(dK)_p) - i\sum_\ell K_{\ell} B_{\star\ell}   +\frac{i}{2\pi}\sum_s \alpha_s\left[ (dB)_{\star s}-2\pi Y_{\star s}\right]\,. \label{eq:anomaly2}
\end{equation}
The fact that we cannot completely remove the gauge variation using background counterterms signals a genuine 't Hooft anomaly. 

Finally, we note that the anomaly persists even if we restrict either $U(1)^{(0)}$ or $U(1)^{(1)}$ to discrete subgroups. For instance, we can restrict the background gauge fields for $U(1)^{(1)}$ to discrete background gauge fields for the $\ZZ_N^{(1)}$ subgroup by setting $B_{\tilde p} = \frac{2\pi}{N}\widehat B_{\tilde p}$, where $\widehat B_{\tilde p} \in \ZZ$, and a discrete gauge redundancy $\widehat B_{\tilde p} \to \widehat B_{\tilde p} + (d\widehat\beta)_{\tilde p} + N L_{\tilde p}$ with $\widehat\beta, L \in \ZZ$. Here $\widehat B_{\tilde p}$ can be regarded as an integer lift of the $\ZZ_N$ gauge field, and should be restricted to be flat modulo $N$, $(d\widehat B)_{\tilde c} \in N\ZZ$.\footnote{This restriction arises dynamically if we Higgs $U(1)^{(1)} \to \ZZ_N^{(1)}$ using a 1-form Stueckelburg field $H_{\tilde\ell} \in \RR$ and action
\begin{equation}
\frac{\lambda}{2}\sum_{\tilde c} \left[ (dB)_{\tilde c} - 2\pi Y_{\tilde c}\right]^2 + \frac{\zeta}{2}\sum_{\tilde p} \left[ (dH)_{\tilde p} - N B_{\tilde p} + 2\pi \widehat B_{\tilde p}\right]^2\,,
\end{equation}
with $\lambda, \zeta \to \infty$ to reach the deep Higgs regime. 
} 

The coupling to the $\ZZ_N^{(1)}$ background field can obtained from Eq.~\eqref{eq:bgdaction1} by setting $B_{\tilde p} = \frac{2\pi}{N}\widehat B_{\tilde p}$ and $Y_{\tilde c} = \frac{1}{N}(d\widehat B)_{\tilde c}$. The anomalous variation becomes
\begin{equation}
\Delta S = \frac{i}{N}\sum_p \widehat \beta_{\star p} \left[ (dA)_p - 2\pi X_p \right] + i \sum_\ell L_{\star\ell}\, A_\ell - \frac{i}{N} \sum_s \alpha_s \left[ (d\widehat B)_{\star s} + N(dL)_{\star s} \right]\,. 
\end{equation}

\section{Worldvolume representations}
\label{sec:worldline_action}

In this Appendix we derive various dual forms of the partition function used in the main text. Our starting point is 
Eq.~\eqref{eq:action_a_chi_pm}, reproduced here for convenience: 
\begin{align}
    S_{\rm gauge} = \frac{\beta}{2} \sum_{p} \> [(da)_{p} - 2\pi m_{p}]^2   +  \frac{\kappa_{\pm}}{2}  \sum_{\pm }\sum_{\ell} \> [ (d 
\varphi_{\pm})_{\ell} \mp \,  a_{\ell} - 2\pi  (n_{\pm})_{\ell} ]^2\,.
\end{align}
We begin by dualizing the gauge field in terms of a sum over worldsheets, and then treat the matter fields. 

\subsection{Dualizing the gauge field}
\label{sec:dual_gauge} 

Dual forms of the action can be obtained directly by applying Poisson summation to the integer-valued fields appearing above. 
Equivalently, we can linearize each of the terms above by introducing real-valued auxiliary variables $\sigma_p$ and $(\rho_\pm)_\ell$. Up 
to an overall constant in the partition function, the above action is dual to 
\begin{align}
\sum_p\left\{ \frac{1}{2\beta}\sigma_p^2 - i  \sigma_p \left[ (da)_p - 2\pi m_p \right] \right\}+ \sum_\pm\sum_\ell  \left\{ 
\frac{1}{2\kappa_\pm}(\rho_\pm)_\ell^2 
- i (\rho_\pm)_\ell \left[ (d\varphi_\pm)_\ell \mp a_\ell - 2\pi (n_\pm)_\ell \right] \right\}\,.
\end{align} 
Summing over $m_p$ constrains $\sigma_p \in \ZZ$. Integrating by parts, the equation of motion for the gauge field $a$ becomes:
\begin{equation}
-(\delta \sigma)_\ell + (\rho_+-\rho_-)_\ell = 0 \,.
\end{equation}
We can solve this for $\rho_-$ to obtain
\begin{align}
    & \sum_p\frac{1}{2\beta}\sigma_p^2
    + \sum_\ell \left\{ \frac{1}{2\kappa_-}(\rho_+ - \delta\sigma)_\ell^2 -i (\rho_+ - \delta\sigma)_\ell 
\left[(d\varphi_-)- 2\pi (n_-)_\ell \right]  \right\}
\nonumber\\ & \qquad\qquad
    + \sum_\ell \left\{ \frac{1}{2\kappa_+}(\rho_+)_\ell^2 -i (\rho_+)_\ell \left[(d\varphi_+)- 2\pi (n_+)_\ell \right]  \right\}
\nonumber\\ & \quad
    =
    \sum_p\frac{1}{2\beta}\sigma_p^2
    + \sum_\ell \left\{ \frac{1}{2\kappa_-}(\rho_+ - \delta\sigma)_\ell^2 + \frac{1}{2\kappa_+}(\rho_+)_\ell^2 -i 
(\rho_+)_\ell \left[ (d\varphi_c) - 2\pi (n_c)_\ell \right] \right\} ,
\end{align} 
where we have defined the `composite' gauge invariant field $\varphi_c \equiv \varphi_+ + \varphi_-$, and $n_c \equiv n_+ + n_-$. We also 
dropped $i\delta\sigma  (d\varphi_--2\pi n_-)$, since the first term is a total derivative and the other is a multiple of $2\pi i$. Note that the 
linear combination $\varphi_+ - \varphi_-$ is absent from the action, and integrating over it contributes an overall infinite constant to the 
partition function which we ignore. Gaussian integration over $\rho_+$ gives
\begin{align}
\label{eq:no_z_plus}
\sum_p \frac{1}{2\beta} \sigma_p^2 + \sum_\ell\left\{ \frac{\kappa_c}{2}[(d\varphi_c)_\ell-2\pi (n_c)_\ell]^2 
+ \frac{\kappa_c}{2\kappa_+\kappa_-}(\delta \sigma)_\ell^2 + 2\pi i \frac{\kappa_c}{\kappa_-}  (n_c)_\ell (\delta\sigma)_\ell \right\} \, .  
\nonumber
\end{align}
where we dropped a total derivative and defined
\begin{align}
    \kappa_c \equiv \frac{\kappa_+ \kappa_-}{\kappa_++\kappa_-}\,.
\end{align}
Finally, the path integral over $\sigma_p \in \ZZ$ can be recast as a sum over collections of surfaces $\Sigma$ by writing $\sigma_p = 
[\Sigma]_p$. These surfaces can have boundaries, which should be interpreted as the worldlines of charged particles. This becomes more 
clear when computing a Wilson loop expectation value using these dual variables. Tracking through the derivation above, a charge-$q$ 
Wilson loop insertion on the contour $C$ modifies the dual form of the action to 
\begin{align} \label{eq:worldsheet_action}
S_{\rm worldsheet}(\Sigma,\varphi_c,n_c; C) &= \frac{1}{2\beta} \sum_p [\Sigma]_p^2 + \frac{\kappa_c}{2}\sum_\ell 
[(d\varphi_c)_\ell-2\pi (n_c)_\ell]^2 \\
&+ \frac{\kappa_c}{2\kappa_+ \kappa_-}\sum_\ell \left([\partial\Sigma]_\ell+q [C]_\ell\right)^2 +2\pi i \frac{\kappa_c}{\kappa_-}\sum_\ell (n_c)_\ell 
\left([\partial\Sigma]_\ell+q [C]_\ell\right)  \,, \nonumber
\end{align}
When $C$ is contractible so that $C = \partial D$, we can perform a shift $\Sigma \to \Sigma - q D$ to absorb the effect of the Wilson loop into the first term. Note that the action is invariant under the gauge symmetries 
\eqref{eq:full_gauge_transformations} of the remaining field variables, 
\begin{equation}
(\varphi_c)_s \to (\varphi_c)_s + 2\pi (k_+ + k_-)_s, \quad (n_c)_\ell \to (n_c)_\ell + 2\pi (dk_+ + dk_-)_\ell\,.
\end{equation}
Furthermore, when $\kappa_+ = \kappa_-$, the action (with $q=0$ above) is invariant mod $2\pi i$ under $\Sigma \to - \Sigma$. 

\subsection{Dualizing the matter fields}
\label{sec:dual_matter}

The above dual action is most useful when $\beta \ll 1$ and $\kappa_c \gg 1$. To analyze the strong gauge coupling, large mass regime 
($\beta 
 \ll 1, \kappa_c \ll 1$), it is convenient to also dualize $(\varphi_c, n_c)$ in terms of composite neutral particle worldlines. To this end, 
we linearize the $d\varphi_c$ kinetic term and sum over $n_c$ to arrive at 
\begin{align}
\sum_p \frac{1}{2\beta}[\Sigma]_p^2+ \sum_\ell\left\{ \frac{1}{2\kappa_c}\left(u_\ell - \frac{\kappa_c}{\kappa_-}[\partial\Sigma]\right)^2+ 
\frac{\kappa_c}{2\kappa_+ \kappa_-}[\partial\Sigma]_\ell^2 \right\}  + i \sum_s (\varphi_c)_s (\delta u)_s\, ,
\end{align}
where $u_\ell \in \ZZ$ is a discrete dynamical field dual to $n_c$. We also need to consider the mixing term
\begin{align}
 \frac{\epsilon}{2} \sum_s 
    \left((\varphi_{0})_s - (\varphi_c)_s - 2\pi t_s \right)^2 \,.
\end{align}
The integral over $\varphi_c$ is Gaussian, and gives
\begin{align}
\sum_p \frac{1}{2\beta}[\Sigma]_p^2 &+ \sum_\ell\left\{ \frac{1}{2\kappa_c}\left(u_\ell - \frac{\kappa_c}{\kappa_-}[\partial\Sigma]\right)^2+ 
\frac{\kappa_c}{2\kappa_+ \kappa_-}[\partial\Sigma]_\ell^2 \right\}  + \sum_s \left\{\frac{1}{2\epsilon}(\delta u)_s^2 + i (\varphi_0)_s(\delta 
u)_s\right\}\, ,
\end{align}
The path integral over the dual variable $u_\ell$ can be recast as the sum over worldlines $\Xi$ of the composite neutral particle 
$\sim 
e^{i\varphi_c}$. Rewriting $u_\ell = [\Xi]_\ell$, we arrive at the final worldvolume form of $S_{\rm gauge} + S_{\rm mixing}$, 
\begin{align} \label{eq:worldvolume_action}
S_{\rm worldvolume}(\Sigma,\Xi, \varphi_0) = \sum_p \frac{1}{2\beta}[\Sigma]_p^2 &+ \sum_\ell\left\{ 
\frac{1}{2\kappa_c}\left([\Xi]_\ell 
- \frac{\kappa_c}{\kappa_-}[\partial\Sigma]\right)^2+ \frac{\kappa_c}{2\kappa_+ \kappa_-}[\partial\Sigma]_\ell^2 \right\} \\
&  + \sum_s \left\{\frac{1}{2\epsilon}[\partial\Xi]_s^2 + i (\varphi_0)_s[\partial\Xi]_s\right\}\, . \nonumber
\end{align}
Starting from this fully dualized form of the action, one can compute correlation functions systematically in an expansion in the small 
parameters $\beta, \kappa_c, \epsilon$. This expansion is effectively a cluster expansion in worldlines and worldsurfaces. The last term 
indicates that in this cluster expansion, the boundaries of the composite particle worldlines are dressed by insertions of $e^{i\varphi_0}$, 
which carry the compensating charge under $\UB$.  

Finally, tracking through the insertion of a Wilson line $W_q(C)$, one finds 
\begin{align} \label{eq:dualized_insertions}
    S_{\rm worldvolume}(\Sigma,\Xi, \varphi_0; C)
    &= \sum_p \frac{1}{2\beta}[\Sigma]_p^2
\nonumber\\
    &+ \sum_\ell\left\{ \frac{1}{2\kappa_c}\left([\Xi]_\ell - \frac{\kappa_c}{\kappa_-}\left([\partial\Sigma]+q[C]_\ell\right) \right)^2
    + \frac{\kappa_c}{2\kappa_+ \kappa_-}\left([\partial\Sigma]+q[C]_\ell\right)^2 \right\}
\nonumber \\
    &  + \sum_s \left\{\frac{1}{2\epsilon}[\partial\Xi]_s^2 + i (\varphi_0)_s[\partial\Xi]_s\right\}\, .
\end{align}

\bibliographystyle{utphys}
\bibliography{small_circle}

\end{document}